\documentclass[sigconf]{acmart}

\AtBeginDocument{%
  }

\copyrightyear{2024}
\acmYear{2024}
\setcopyright{rightsretained}
\acmConference[SIGSPATIAL '24]{The 32nd ACM International Conference on Advances in Geographic Information Systems}{October 29-November 1, 2024}{Atlanta, GA, USA}
\acmBooktitle{The 32nd ACM International Conference on Advances in Geographic Information Systems (SIGSPATIAL '24), October 29-November 1, 2024, Atlanta, GA, USA}
\acmDOI{10.1145/3678717.3691218}
\acmISBN{979-8-4007-1107-7/24/10}

\acmSubmissionID{25}

\usepackage[ruled]{algorithm2e}
\usepackage{subcaption}

\begin{document}

\title{Critical Features Tracking on Triangulated Irregular Networks by a Scale-Space Method}

\author{Haoan Feng}
\email{hfengac@umd.edu}
\orcid{0000-0002-3667-3990}
\affiliation{%
  \institution{University of Maryland, College Park}
  \state{Maryland}
  \country{United States}
}

\author{Yunting Song}
\email{ytsong@umd.edu}
\orcid{0000-0002-3053-1748}
\affiliation{%
  \institution{University of Maryland, College Park}
  \state{Maryland}
  \country{United States}
}

\author{Leila De Floriani}
\email{deflo@umd.edu}
\orcid{0000-0002-1361-2888}
\affiliation{%
  \institution{University of Maryland, College Park}
  \state{Maryland}
  \country{United States}
}

\renewcommand{\shortauthors}{Feng, et al.}

\begin{abstract}
    The scale-space method is a well-established framework that constructs a hierarchical representation of an input signal and facilitates coarse-to-fine visual reasoning. Considering the terrain elevation function as the input signal, the scale-space method can identify and track significant topographic features across different scales. The number of scales a feature persists, called its \textit{life span}, indicates the importance of that feature. In this way, important topographic features of a landscape can be selected, which are useful for many applications, including cartography, nautical charting, and land-use planning. The scale-space methods developed for terrain data use gridded Digital Elevation Models (DEMs) to represent the terrain. However, gridded DEMs lack the flexibility to adapt to the irregular distribution of input data and the varied topological complexity of different regions. Instead, Triangulated Irregular Networks (TINs) can be directly generated from irregularly distributed point clouds and accurately preserve important features. In this work, we introduce a novel scale-space analysis pipeline for TINs, addressing the multiple challenges in extending grid-based scale-space methods to TINs. Our pipeline can efficiently identify and track topologically important features on TINs. Moreover, it is capable of analyzing terrains with irregular boundaries, which poses challenges for grid-based methods. Comprehensive experiments show that, compared to grid-based methods, our TIN-based pipeline is more efficient, accurate, and has better resolution robustness.
\end{abstract}

\ccsdesc[500]{Information systems~Geographic information systems}
\ccsdesc[500]{Computing methodologies~Mesh geometry models}
\ccsdesc[500]{Computing methodologies~Shape analysis}
\ccsdesc[500]{Computing methodologies~Concurrent algorithms}

\keywords{Scale-space data analysis, Triangulated Irregular Networks, Topological methods, Cartography}


\maketitle

\section{Introduction}\label{sec:intro}
Scale-space methods have been widely utilized in the fields of computer vision and image processing since the pioneering work of Koenderink~\cite{Koenderink1984} and Witkin~\cite{Witkin1987}. The scale-space framework mimics the multi-resolution property of the human visual cortex by representing objects at different scales. The level of detail (LOD) of the input signal decreases as the scale increases. Lindeberg~\cite{lindeberg2013scale} introduced a method to analyze the change of input signal across different scales by tracking the zero-crossings of the differential invariants of the signal. These zero-crossing points, named \textit{critical points} or \textit{features}, describe the topological structure of the input signal. Critical points that persist across scales are particularly essential as they are resistant to the change in scale and represent more prominent data points to convey the topological information of the input signal. 

Features extracted through scale-space methods are valuable for analysis tasks in remote sensing and Geographical Information Systems (GIS)~\cite{ai2010lifespan, kong2016extraction, tie2016extraction, lai2019improving}.  For instance, generating nautical charts from sounding datasets heavily relies on the manual selection of soundings by human experts, who face constraints from safety requirements and cartographic criteria. To support human experts, many analysis systems aim to automatically identify important data points or provide useful auxiliary information~\cite{Dyer2022, Li2021a, Skopeliti2020}. Similarly, critical points can also assist in identifying topologically significant locations (e.g., \textit{Spot heights}) on a map that potentially satisfies the cartographers' requirements.

When using scale-space methods to extract critical features from terrain, gridded terrain surface models, i.e., gridded Digital Elevation Models (gridded DEMs) are usually used~\cite{Rocca2017}. However, when the input point clouds are distributed irregularly, the process of interpolating them into regular grid cells causes a loss of information. Besides, the fixed cell size of gridded DEMs poses difficulty in capturing the irregularity of the terrain surface with minimal costs. For instance, a coarser grid reduces the storage cost but cannot fully represent the details in mountainous regions, while a finer grid requires a higher storage cost.  Therefore, when input points are distributed irregularly, Triangulated Irregular Networks (TINs) are usually used to represent the terrain surface instead.  TINs are more efficient in representing topographical information at multiple resolutions for different regions of various complexity. 
Moreover, when represented as a gridded DEM, the boundary information of the terrain surface may be lost or roughly approximated. Thus, in applications where boundary information is crucial, such as coastal and ocean modeling~\cite{Candy2014, Avdis2018}, TINs are preferred as they can better represent line features than gridded DEMs.  

In this work, we present a novel pipeline for scale-space tracking of critical points on TINs inspired by the scale-space method defined for gridded DEMs in~\cite{Rocca2017}. Our experiments show that direct analysis of a TIN through the scale-space method can produce a more accurate result with limited storage. Moreover, we address several challenges, in performing a scale-space method on TINs due to the irregularity of TINs. Specifically, the main contributions of this work are:

\begin{itemize}
    \item   A strategy to generate high-quality TINs for scale-space analysis: we propose to employ a local curvature-based sampling method and $\alpha$-shape method~\cite{akkiraju1995alpha} to generate TINs from raw point sets. Experiments show that, compared to the gridded DEM, the resulting TIN preserves more surface details and is more robust when available storage is limited.
    \item   Critical features tracking on TINs: we extend the previous tracking method~\cite{Rocca2017} for gridded DEMs to TINs with special consideration of varied configurations of adjacent vertices and complex surface topology involving $k$-fold saddles.
    \item   TIN smoothing: scale-space methods require a smoothing process that satisfies the heat-diffusion model. However, a direct extension from image smoothing to TINs leads to inaccurate results. To bridge this gap, we propose a novel iterative Gaussian filtering method customized for TIN smoothing.
    \item   Parallelization: we design and implement a parallel TIN smoothing procedure taking advantage of modern GPU architecture to significantly speed up the smoothing process. Additionally, the scale-space analysis stage is also parallelized with the help of the Terrain trees~\cite{fellegara2023terrain}.
\end{itemize}

The remainder of the paper is organized as follows. Section~\ref{sec:bg} provides background information and the problem setup, followed by Section~\ref{sec:rel}, which reviews previous work on scale space.  
In Section~\ref{sec:overview}, we review the discrete grid-based scale-space framework for gridded terrains proposed in~\cite{Rocca2013,Rocca2017}. Section~\ref{sec:method} presents our pipeline for scale-space critical point tracking method on TINs. In Section~\ref{sec:exp}, we present an experimental comparison of our TIN-based method to the previous grid-based method. Finally, in Section~\ref{sec:discuss}, we discuss a potential application to bathymetric data and summarize key innovative points and future development of our pipeline.

\section{Background}\label{sec:bg}
TINs are defined based on discrete point sets in the plane, where each data point has an elevation associated with it.  A TIN consists of the set of vertices $\mathcal{V}$ and connectivity information $\mathcal{C}$ describing how the vertices are connected through the triangle mesh. Each vertex $v$ in a TIN consists of a \textit{Point} $p \in \mathbb{R}^2$ representing the $xy$-coordinate, and an elevation value. Besides vertices, triangles $\mathcal{T}$ are the primary components of a TIN, from which the connectivity information $\mathcal{C}$ can be derived.

\begin{figure}[t]
    \centering
    \includegraphics[width=0.8\linewidth]{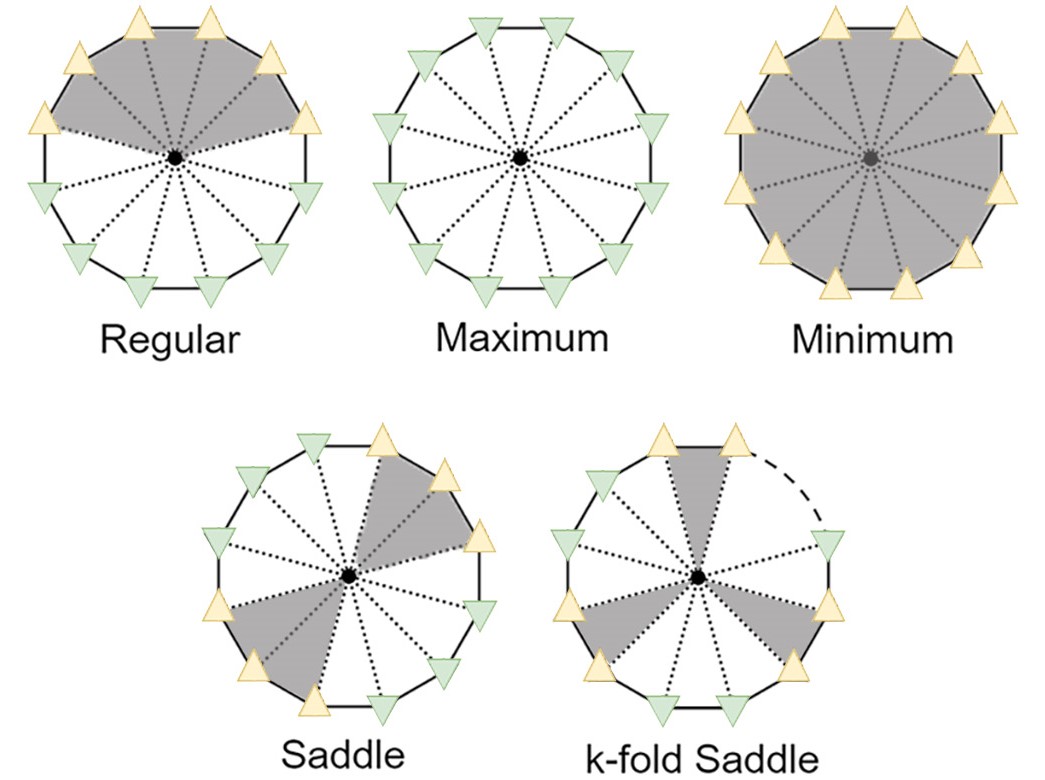}
    \caption{Critical point types by \textit{vertex signature}. For each neighbor vertex, {\color{yellow}$\blacktriangle$} indicates its elevation is higher than the center vertex; otherwise, {\color{green}$\blacktriangledown$} is drawn. Derived from~\cite{edelsbrunner2001hierarchical}.\label{fig:point_type}}
    \Description{The figure shows a center vertex surrounded by six neighboring vertices. The center vertex is marked with a black circle. The neighboring vertices are marked with black triangles or black triangles pointing downwards. The black triangles indicate that the neighboring vertex has a higher value than the center vertex. The black triangles pointing downwards indicate that the neighboring vertex has a lower value than the center vertex. The center vertex is classified as a maximum, minimum, or saddle based on the number of neighboring vertices with higher or lower values.}
\end{figure}

Critical features, such as peaks, saddles, and pits, on a TIN can be extracted by using Banchoff's theory~\cite{banchoff1970critical} which defines critical points for polyhedral surfaces and is one of the pillars of Piece-wise Linear (PL) Morse theory\cite{edelsbrunner2001hierarchical}. According to~\cite{banchoff1970critical}, a vertex $v$ in a TIN $\Theta$ can be classified into one of five categories: \textit{regular}, \textit{maximum}, \textit{minimum}, \textit{saddle}, and $k$\textit{-fold saddle}, as shown in Figure~\ref{fig:point_type}. The classification is done through a comparison between elevations at $v$ and its adjacent vertices $[v_1, v_2, \ldots,v_m]$ listed clockwise. To this aim, the \textit{vertex signature} is defined as the sequence of Boolean values $[v>v_1, v>v_2, \ldots, v>v_m]$. The head and tail of this sequence are considered neighbors in this list. As annotated in shadow in Figure~\ref{fig:point_type}, consecutive Booleans with the same value are grouped into \textit{higher/lower components}, denoting neighboring vertices belonging to them having higher/lower elevations than $v$ does. For instance, if the vertex signature of $v$ is [\textit{true, false, true, false, true}], the number of higher/lower components $k_{higher}/k_{lower}$ equals $2$ since the last and first vertices that are \textit{true} are adjacent and belong to the same component. It helps determine the point type of $v$. In general, a vertex $v$ is a \textit{maximum/minimum} when $k_{higher} = 0/k_{lower} = 0$. And, a $k$-fold saddle has $k_{higher}=k_{lower}=k+1$. When $k = 1$, $v$ is considered of type \textit{simple saddle}.

The Poincar\'{e}-Hopf index theorem~\cite{milnor1963morse} specifies that the numbers of maxima, saddles, and minima on a polyhedral surface (and thus on a TIN) follow an invariant relation:
\begin{equation}\label{equa:index}
    N_{max} + N_{min} - N_{saddle} = 2
\end{equation}
In a scale space, this invariant property puts a constraint on the number of critical points and their interactions on a polyhedral surface at any scale. It helps to understand how critical points interact and transition, which makes the tracking of critical points across scales possible.

\section{Related Work}\label{sec:rel}
Scale-space methods were first systematically introduced by Koenderink~\cite{Koenderink1984} and Witkin~\cite{Witkin1987}, and have since been extensively studied in various fields such as computer vision, image processing, and signal analysis. A Scale space is generally defined as a single-parameter family $f(s)$ of a signal $f$, where the scale $s$ is the only parameter. Imitating the human visual system, the captured visual signal contains various features that become prominent at different scales. Invariant features are considered more important in the visual recognition process as the scale progresses from fine to coarse. Scale-space methods process the input signal $f$ iteratively using filters that satisfy the heat-diffusion model, such as the Gaussian kernel. As the scale increases, the kernel size $\sigma$ increases correspondingly. This process progressively eliminates finer details of the signal while retaining a clue from the coarse features to some finer details that have been removed, leaving only the coarse features at each scale layer. By sharing the same idea with the \textit{Deep Structure} reasoning proposed by Lindeberg~\cite{lindeberg2013scale}, the topological importance of critical points of the function $f$ can be obtained by tracking their interactions and analyzing their persistence across scales.

On gridded DEMs, critical point identification and cross-scale correspondence matching are well explored. Lin et al.~\cite{tie2016extraction} utilized scale space to filter out unnecessary finer details for ridge line detection and merge the ridge line scratches across scales for improved results. Li et al.~\cite{Li2021a} utilized topological features from bathymetric sounding data to aid in the production of nautical charts. They identified feature points on each scale and used a proximity-based matching approach between consecutive scales, known as \textit{neighbor searching}, to find critical feature points that exist across multiple scales. However, this matching approach can be error-prone, especially when the scale is rough and critical point movement is substantial~\cite{Rocca2017}. To improve the neighbor searching of critical points' correspondence, Reininghaus et al.~\cite{reininghaus2011scale} considered the underlying gradient of the scalar function. This change improves the robustness to noise and the matching rate of the neighbor searching. However, there still exists uncertainty in the matching of maxima and minima at rough scales.

\begin{figure}[t]
    \centering
    \includegraphics[width=0.85\linewidth]{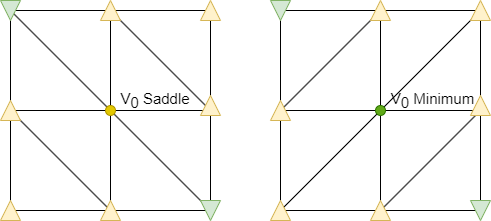}
    \caption{Preset edge connection of a gridded DEM affects critical point types. Vertices with {\color{yellow}$\blacktriangle$}/{\color{green}$\blacktriangledown$} markers have elevations higher/lower than $v_0$. For the type of $v_0$, the left model implies \textit{Saddle}, while the right model implies \textit{Minimum}.}
    \label{fig:wrongtype}
    \Description{The figure shows two different edge connection models around a center vertex. The left connection model: the center vertex has edges connecting its horizontal and vertical neighbors plus the NW and SE diagonal neighbors. The right connection model: the center vertex has edges connecting its horizontal and vertical neighbors plus the NE and SW diagonal neighbors. Height values of the neighboring vertices: the horizontal, vertical, NE, and SW neighbors have higher values, and NW and SE diagonal neighbors have lower values.}
\end{figure}

Rocca et al.~\cite{Rocca2013, Rocca2017} proposed a more robust method for tracking the transformation of critical points for gridded DEMs. Their method defines a virtually continuous scale space where the movements and interactions of critical points across scales can be explicitly tracked. However, this method works only for gridded datasets. A local regular triangle mesh is generated from the grid by splitting each grid cell into two through a diagonal to identify the critical points. This choice of the diagonal leads to different critical point labeling results as illustrated in Figure~\ref{fig:wrongtype}.

In this paper, we extend the \textit{grid-based method} proposed by Rocca et al.~\cite{Rocca2017} to TINs, which are defined by irregular meshes.  A discrete scale space based on a TIN is constructed for critical features tracking from raw point clouds. This extension is not trivial due to the different spatial subdivisions used by TINs and gridded DEMs and to the irregularity of TINs. 

\section{A discrete scale-space method on regular triangle meshes}\label{sec:overview}
In~\cite{Rocca2017}, a discrete scale space $\Phi$ on a gridded DEM is constructed by iteratively processing the input gridded DEM $f$ with a Gaussian smoothing operator $g$, i.e., $\Phi = \{f, g(f), g(g(f)), ...\}$. They convert the input gridded DEM into a regular triangle mesh by splitting each grid cell into two triangles through one diagonal as cases illustrated in Figure~\ref{fig:wrongtype}. Each element of the scale space is a regular triangle mesh with the same vertices and triangles as others, except vertex elevations are different, representing the input $f$ at different scales. From fine to coarse, the elements in $\Phi$ are denoted as layers at different scales $s_i$ for $i \in \mathbb{N}$, i.e., $\Phi = \{s_0, s_1, s_2,...\}$. The discrete scale space $\Phi$ is defined as a piecewise-linear space based on these layers. Thus, the intermediate scale space between two consecutive layers $s_i$ and $s_{i+1}$ is obtained by linear interpolation. Thereby, the intermediate scale $s_{i+\delta}$ can be calculated as 
\begin{equation}
s_{i+\delta} = \delta s_{i + 1} + (1 - \delta) s_{i}
\end{equation}
where $0 < \delta < 1$. From this equation, the elevation of each vertex $v$ is determined in the scale space as $s_{t}(v)$, where $t$ represents the scale of the input $f$, also known as \textit{timestamp} in this context.
 
As detailed in Section~\ref{sec:bg}, the point type of a vertex $v_m$ depends only on the relationship to its adjacent vertices in the mesh in terms of relative elevation. Thus, the type of $v_m$ can only change when its relative elevation to one of its neighbors changes. Suppose this neighbor vertex is $v_n$ and this change happens at timestamp $t$, then these two vertices have the same elevation value at timestamp $t$, i.e., $s_t(v_m) = s_t(v_n)$. Formally, we have:
\begin{equation}
Sign\left(s_{t-\epsilon}(v_m) - s_{t-\epsilon}(v_n)\right) = -Sign\left(s_{t+\epsilon}(v_m) - s_{t+\epsilon}(v_n)\right)
\end{equation}
where $\epsilon$ represents a small change in the timestamp. The values at the extremes of the edge between $v_m$ and $v_n$ swap. We call this phenomenon an \textit{edge flip} at timestamp $t$, where $t \in [i, i+1)$. Following the piecewise-linear assumption of this discrete scale space, this timestamp $t_\mathrm{edge\ flip}$ is calculated as:
\begin{equation}
    t_\mathrm{edge\ flip} = i + \frac{s_i(v_m) - s_i(v_n)}{s_{i+1}(v_n) - s_i(v_n) + s_i(v_m) - s_{i + 1}(v_m)}
\end{equation}

Since the type of a vertex changes only during an edge flip, tracking the critical points in the discrete scale space $\Phi$ is equivalent to processing edge-flip events in chronological order.

\begin{figure}[h]
    \centering
    \includegraphics[width=0.9\linewidth]{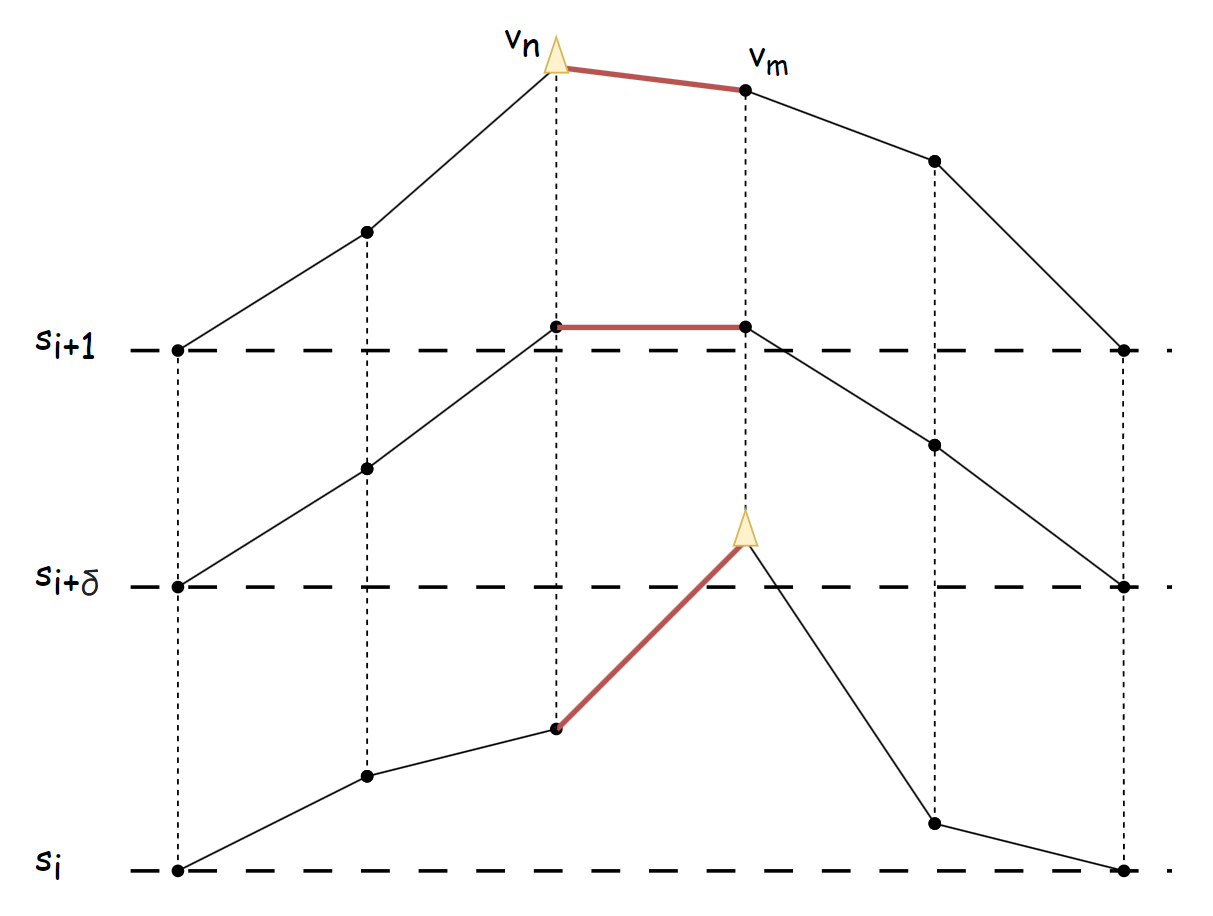}
    \caption{A 1D illustration of maximum transition at the edge-flip event. {\color{yellow}$\blacktriangle$} denotes the maximum moving from vertex $v_m$ to $v_n$. Intermediate scale $s_{i+\delta}$ illustrates the timestamp that the edge flip happens.}
    \Description{This figure draws a 1D slightly screwed bell-shaped signal in the scale space. From one scale layer to another, the peak shifts from one vertex to its adjacent vertex. The intermediate edge flip timestamp is also drawn, in which the edge connecting these two vertices is flat.}
    \label{fig:cp_dis}
\end{figure}

To better illustrate this process, we take the critical point transition of a 1D signal as an example (see Figure~\ref{fig:cp_dis}). Vertex $v_m$ is a maximum at the scale $s_i$, and it becomes a regular point at the scale $s_{i+1}$ as one of its adjacent vertices now has a higher value than itself. Meanwhile, vertex $v_n$ becomes a maximum at the scale $s_{i+1}$ since it now has a higher value than all adjacent vertices. These changes happen because the edge $\overline{v_mv_n}$ flips at timestamp $i+\delta$ and the relative relation between $v_m$ and $v_n$ is changed. Thus, an edge flip happens at timestamp $i+\delta$ makes a maximum move from vertex $v_m$ to $v_n$.

In the \textit{grid-based method} by~\cite{Rocca2017}, the transition of critical points is limited to preset cases because the edge connection is fixed and each internal vertex of the mesh has six neighbors. However, on a TIN, the number of vertex neighbors is not constant and the existence of $k$-fold saddles makes the critical point transition much more complex than the preset cases in the grid-based method.

\section{Methodology}\label{sec:method}
We propose a TIN-based pipeline for feature tracking through a scale-space method. The proposed pipeline takes raw point cloud data as input and outputs tracked critical points with a quantitative measure of their importance. Point cloud data are obtained from various sources such as Light Detection and Ranging (LiDAR) scans of landscape terrain~\cite{reutebuch2005light}. The pipeline is composed of three consecutive stages: (I) input point cloud preprocessing (see Section~\ref{sec:preprocess}), (II) TIN generation and encoding (see Section~\ref{sec:tingen}), and (III) scale-space analysis on a TIN (see Section~\ref{sec:scalespace}).

\subsection{Point cloud preprocessing}\label{sec:preprocess}
In the scale-space analysis, critical points move and interact with each other through the evolution from fine to rough scales. In the discrete setting, the trajectory of a critical point is approximated by a continuous path of TIN edges. Thus, the quality of the TIN directly affects the accuracy of the critical point tracking result. However, terrain point clouds are irregularly distributed and contain outliers, leading to erroneous TIN smoothing results and a waste of computational resources. It is necessary to eliminate outliers and prevent data point overlap. 
One widely used method to down-sample point cloud is the Furthest Point Sampling (FPS) method~\cite{eldar1997farthest}. However, the FPS method leads to an approximately uniform density of vertices across the terrain surface. This is not ideal for representing the terrain surface with varying complexity, since complex landscapes have more details and should be described with more vertices.

\begin{figure}[t]
    \centering
    \begin{subfigure}{0.49\columnwidth}
        \centering
        \includegraphics[width=0.95\linewidth]{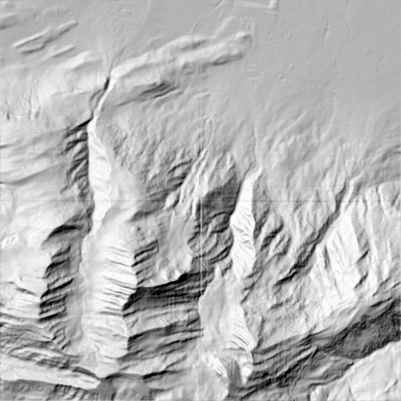}
        \caption{Hillshade}
    \end{subfigure}
    \begin{subfigure}{0.49\columnwidth}
        \centering
        \includegraphics[width=0.95\linewidth]{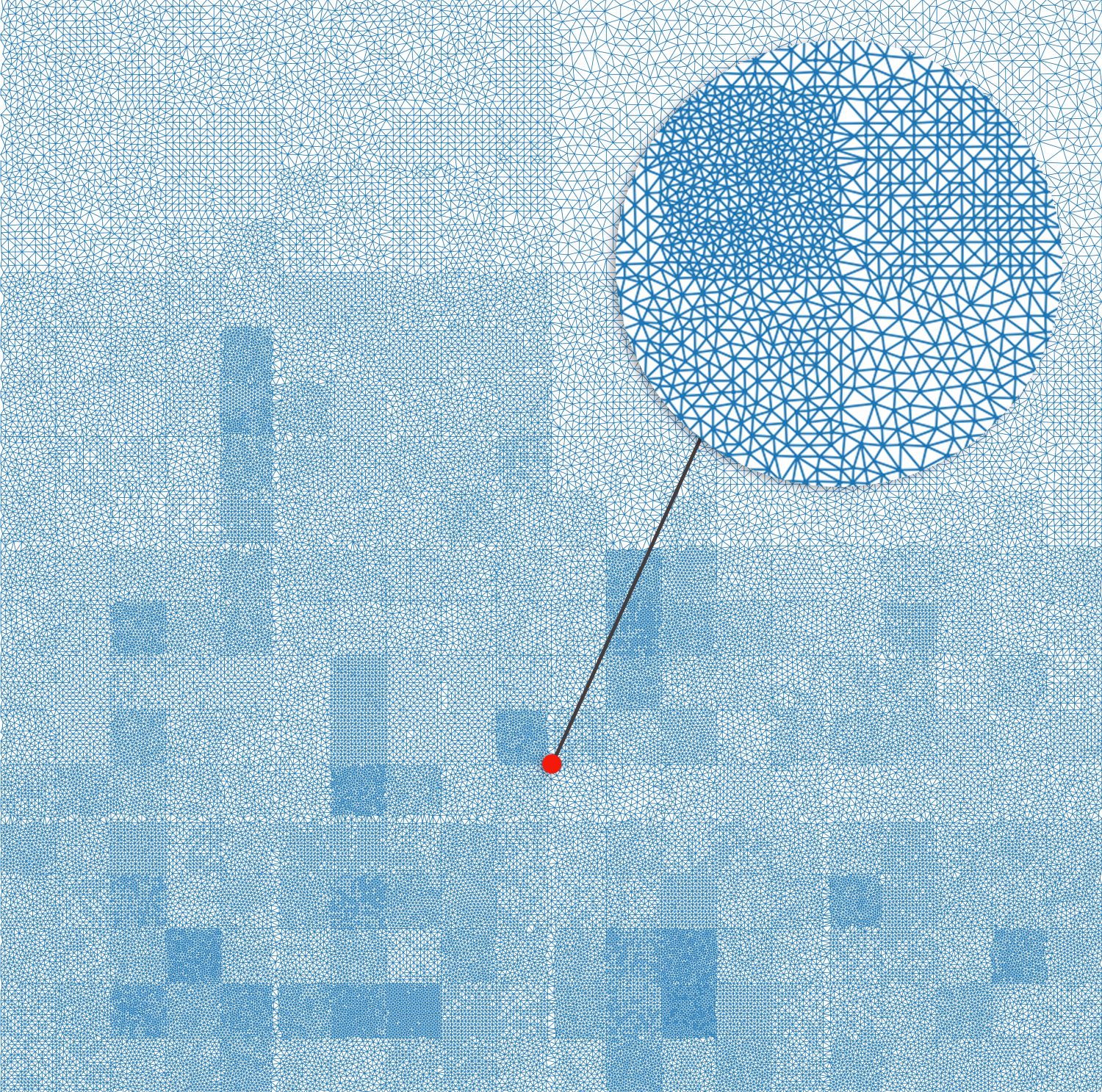}
        \caption{TIN wireframe}
    \end{subfigure}
    \newline
    \begin{subfigure}{0.49\columnwidth}
        \centering
        \includegraphics[width=0.95\linewidth]{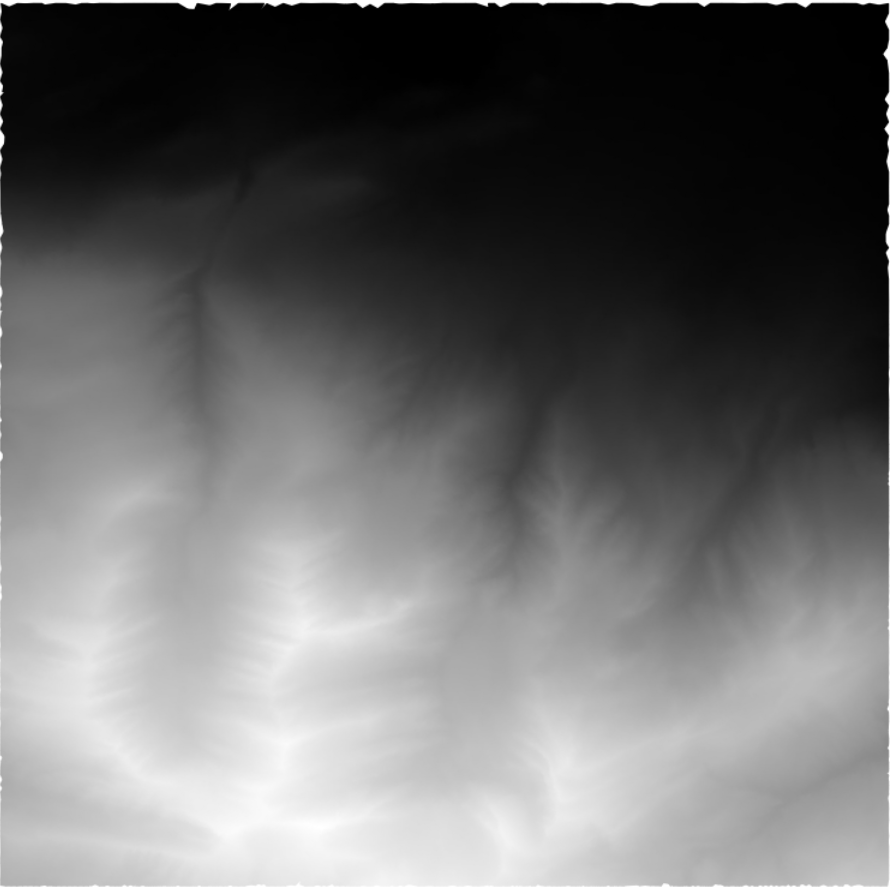}
        \caption{TIN BEV}
    \end{subfigure}
    \begin{subfigure}{0.49\columnwidth}
        \centering
        \includegraphics[width=0.95\linewidth]{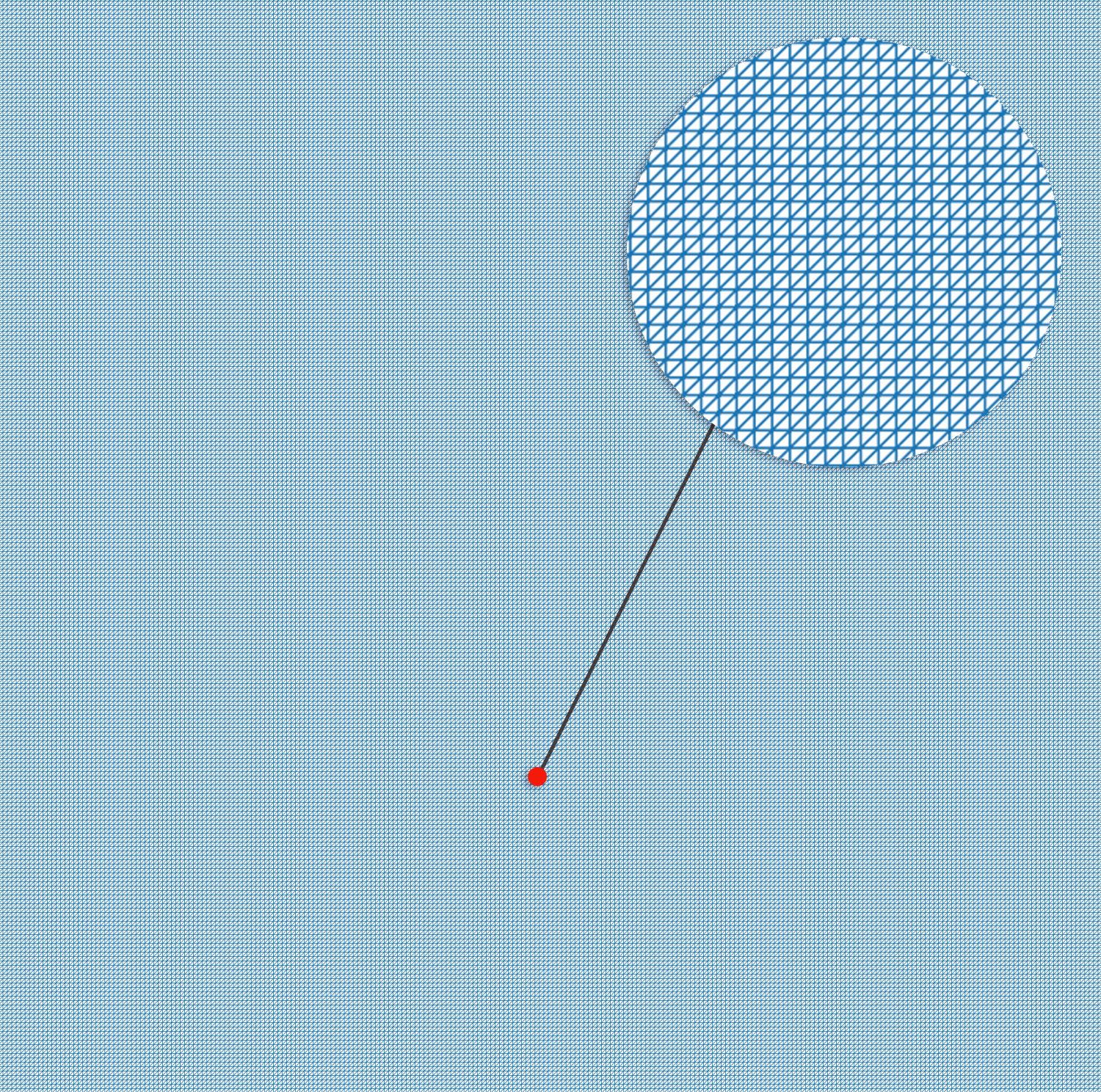}
        \caption{Gridded DEM wireframe}
    \end{subfigure}
    \caption{PFPS method result of Reichenburg suburban region. (a) is the hillshade of the Reichenburg Suburban region, and (c) is the Bird's Eye View (BEV) of the TIN generated. (b) and (d) are wireframe plots of the TIN and gridded DEM.}
    \label{fig:visualsampling}
    \Description{The figure shows the hillshade of the Reichenburg suburban region, the Bird's Eye View (BEV) of the TIN generated, and the wireframe plots of the TIN and gridded DEM. In the suburban region, TIN has a sparser mesh while the gridded DEM has the same cell size everywhere.}
\end{figure}

To address this issue, we propose a Patch-based extension of the FPS method (PFPS), which divides the terrain into equal-sized regions and samples varying numbers of points from different regions of the terrain surface based on the per-patch estimated curvature. This curvature is defined as the ratio of the largest eigenvalue of the covariance matrix to the sum of all eigenvalues. This value represents the local terrain complexity and a patch with a small curvature tends to be a flat region. Based on the estimated curvature, the number of points to be sampled from each patch can be determined, and within each patch, the FPS method is applied. As a result, in Figure~\ref{fig:visualsampling}(b), patches belonging to the flat suburban area in the terrain dataset require fewer points to be represented than mountainous regions need.

 The PFPS algorithm is presented in Algorithm~\ref{alg:pfps} and a comparison between different down-sampling methods is shown in Appendix~\ref{appendix:downsampling}.

\begin{algorithm}
    \caption{Patch-based Furthest Point Sampling}\label{alg:pfps}
    \KwIn{Point cloud $pcd$, patch size $s$, percentage of points to keep $\rho$}
    \KwOut{Down-sampled point cloud $pcd_{down}$}
    $pcd_{down} \gets \emptyset$\\
    // {\color{blue} Divide the point cloud into patches over $xy$-plane}\\
    $patches \gets subdivide(pcd, s)$\\
    \ForEach{$patch$ in $patches$}{
        // {\color{blue} Estimate the local curvature via the covariance matrix}\\
        $covmat \gets Covariance~Matrix~(patch)$\\
        $\lambda_1, \lambda_2, \lambda_3 \gets Eigenvalues~(covmat)$\\
        // {\color{blue} $\lambda_1 \ge \lambda_2 \ge \lambda_3$}\\
        $patch_{curv} \gets \lambda_1 /(\lambda_1 + \lambda_2 + \lambda_3)$\\
    }
    // {\color{blue} Normalize the estimated curvatures as sampling weights}\\
    $patch_{w} \gets patch_{curv} / \sum_{patches} patch_{curv}$\\
    \ForEach{$patch$ in $patches$}{
        $pcd_{patch} \gets FPS~(patch, \rho \times patch_{w})$\\
        $pcd_{down} \gets pcd_{down} \cup pcd_{patch}$\\
    }
    \KwRet{$pcd_{down}$}
\end{algorithm}

\subsection{TIN generation and encoding}\label{sec:tingen}
\begin{figure}[!htbp]
    \centering
    \includegraphics[width=\linewidth]{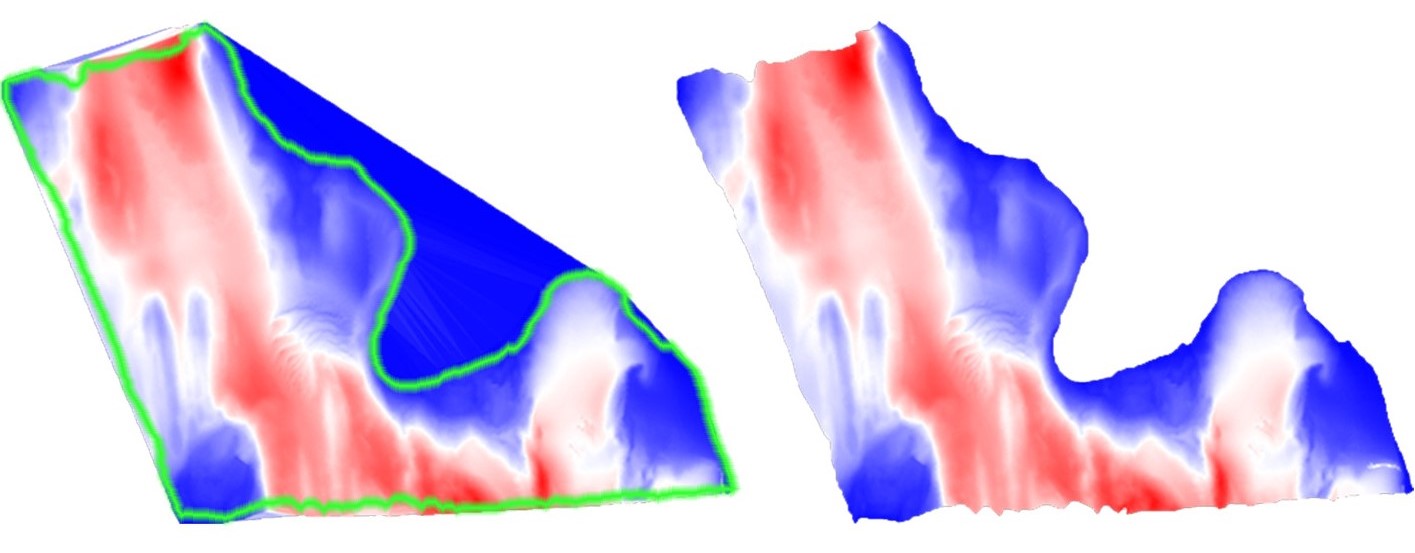}
    \caption{$\alpha$-shape method (right) removes long triangles on the border from the Delaunay triangulation result (left). Green lines in the left figure highlight the boundary of the processed triangle mesh.}\label{fig:alphashape}
    \Description{For a BEV of bathymetry dataset, Delaunay triangulation result (left) and the result after applying the $\alpha$\textit{-shape} method(right) in which the flat triangles on the border of the domain are removed. Green lines in the left figure highlight the boundary of the processed triangle mesh.}
\end{figure}

As introduced in Section~\ref{sec:bg}, a TIN consists of a 2D triangle mesh and a scalar function (elevation) defined on it. The underlying triangle mesh of a TIN is generated through a Delaunay triangulation~\cite{fortune2017voronoi} of the preprocessed point cloud. In our pipeline, the transitions of critical points are tracked along the edges of the TIN, as detailed in Section~\ref{sec:tracking}. Thus, sliver triangles may lead to erroneous results. For example, in Figure~\ref{fig:alphashape}, the long edges on the boundary of TIN should be removed to avoid critical points accidentally transiting through them.

To address this issue, we apply the $\alpha$\textit{-shape} method~\cite{akkiraju1995alpha} to the underlying triangle mesh of the TIN, which defines a filter criterion to remove triangles with circumradii larger than a preset threshold $\alpha$. The $\alpha$-shape method removes long triangles on the border of the TIN, which also reduces the time and memory costs for the following scale-space analysis. Furthermore, this step is especially beneficial for datasets with irregular boundaries, like the bathymetry dataset shown in Figure~\ref{fig:alphashape}.

One challenge that hinders the usage of TINs for large-scale terrain analysis is the high memory and time costs of processing TINs. In this work, we use a compact data representation for TINs, the Terrain trees~\cite{fellegara2023terrain}, to encode TINs and to perform analysis operations. Terrain trees surpass commonly utilized structures like the Indexed data structure with Adjacencies (IA data structure) regarding memory usage and local neighbor extraction performance~\cite{fellegara2023terrain}. Leveraging Terrain trees, we are able to process large-scale TINs with less memory and latency. Moreover, the spatial decomposition used by Terrain trees also allows for parallel processing using OpenMP~\cite{chandra2001parallel}, further enhancing the computation speed and efficiency.

\subsection{Virtually continuous scale-space analysis on a TIN}\label{sec:scalespace}
Inspired by virtually continuous scale-space by~\cite{Rocca2013}, we propose a discrete scale-space method based on TINs, which consists of two parts: the construction of scale space from the generated TIN by a sequence of smoothing operations (see Section ~\ref{sec:smoothing}) and the tracking of critical points across scales (see Section~\ref{sec:tracking}).

\subsubsection{Iterative TIN smoothing}\label{sec:smoothing}
As mentioned in Section~\ref{sec:overview}, the scale space of a gridded DEM is generated through a sequence of Gaussian smoothing operations. Gaussian smoothing~\cite{gonzalez2009digital} is well-defined in the image domain and efficiently achieved through convolution operations. The Gaussian smoothing kernel size, denoted as $\sigma$, increases as the scale increases. Iteratively processing the smoothed layer by a larger Gaussian kernel, the scale space is established by stacking these smoothing results. Heuristically, the kernel size $\sigma_i$ for the $i$-th layer of the scale space is set as $\sigma^2_i = 2^i$. Applying smoothing to a gridded DEM is straightforward, with convolution matrices' size linearly increasing with the kernel size. For a TIN, however, this process requires more adaptation to account for the irregular spatial decomposition inherent in a TIN.

To bridge this gap, we introduce an iterative approximation of the Gaussian smoothing process on a TIN $\Theta$ using a small, fixed kernel size, leveraging the semi-group property of the Gaussian kernel~\cite{Delio2007}. This property states that consecutively smoothing operations with two kernels of sizes $\sigma_1$ and $\sigma_2$ are equivalent to a single smoothing operation with a kernel of size $\sqrt{\sigma_1^2 + \sigma_2^2}$. By presetting a small kernel size $\sigma_{small}$, we limit the local influence region of a single smoothing operation. The small kernel size leads to more iterations and longer computation time. Therefore, we utilize parallel computing on GPU to accelerate the smoothing process. The smoothing process can be regarded as the weighted average of adjacent vertices. It can be calculated through a matrix-vector multiplication 
\begin{equation}
    \vec{f}^{(t)} = M_{w}\vec{f}^{(t-1)}
\end{equation}
where vector $\vec{f}^{(t)}$ represents the scalar function values of TIN vertices at timestamp $t$ and the weight matrix $M_{w}$ represents the weights of each vertex's adjacent vertices. $M_{w}$ is a $V \times V$ matrix, where $V$ is the number of vertices in $\Theta$ and the matrix element at $(i, j)$ is the weight of edge $\overline{v_iv_j}$ connecting vertices $v_i$ and $v_j$. To fit into the GPU memory, a sparse representation of $M_{w}$ is employed. With the fixed kernel size $\sigma_{small}$, we compute the weight matrix $M_{w}$ as follows:
\begin{equation}\label{equa:weight}
    M_{w}[i, j] =
    \begin{cases}
        \frac{1}{Z} \exp\left(-d(i, j)^2 / \left(2\sigma_{small}^2\right)\right) & \text{if } j \in Neighbor(i) \\
        0                                                                   & \text{otherwise}
    \end{cases}
\end{equation}
where $d(i, j)$ is the distance between $v_i$ and $v_j$ on the $xy$-plane, and $Z$ is the normalization factor that ensures the total weight of vertex $v_i$ sums to one. The smoothing process is then executed by performing sparse matrix-vector multiplication on GPUs as:
\begin{equation}
    \vec{L}_{t + 1} = sparse\_matvec(M_{w}^{t},\vec{L}_{t})
\end{equation}
where $L_t$ represents the scalar function value of the $t$-th layer of the scale space in vector form and $M_{w}^{t}$ is the transition matrix.

\begin{figure}[t]
    \centering
    \begin{subfigure}{0.49\columnwidth}
        \centering
        \includegraphics[width=\textwidth]{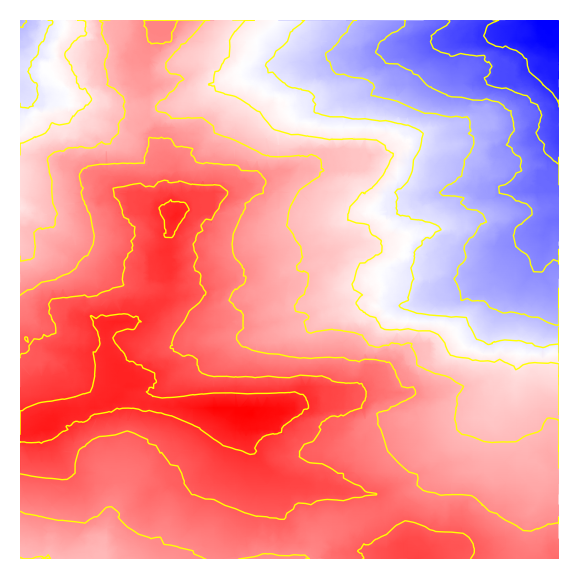}
        \subcaption{TIN smoothing w.o. virtual neighbors and angle re-weighting}
    \end{subfigure}
    \begin{subfigure}{0.49\columnwidth}
        \centering
        \includegraphics[width=\textwidth]{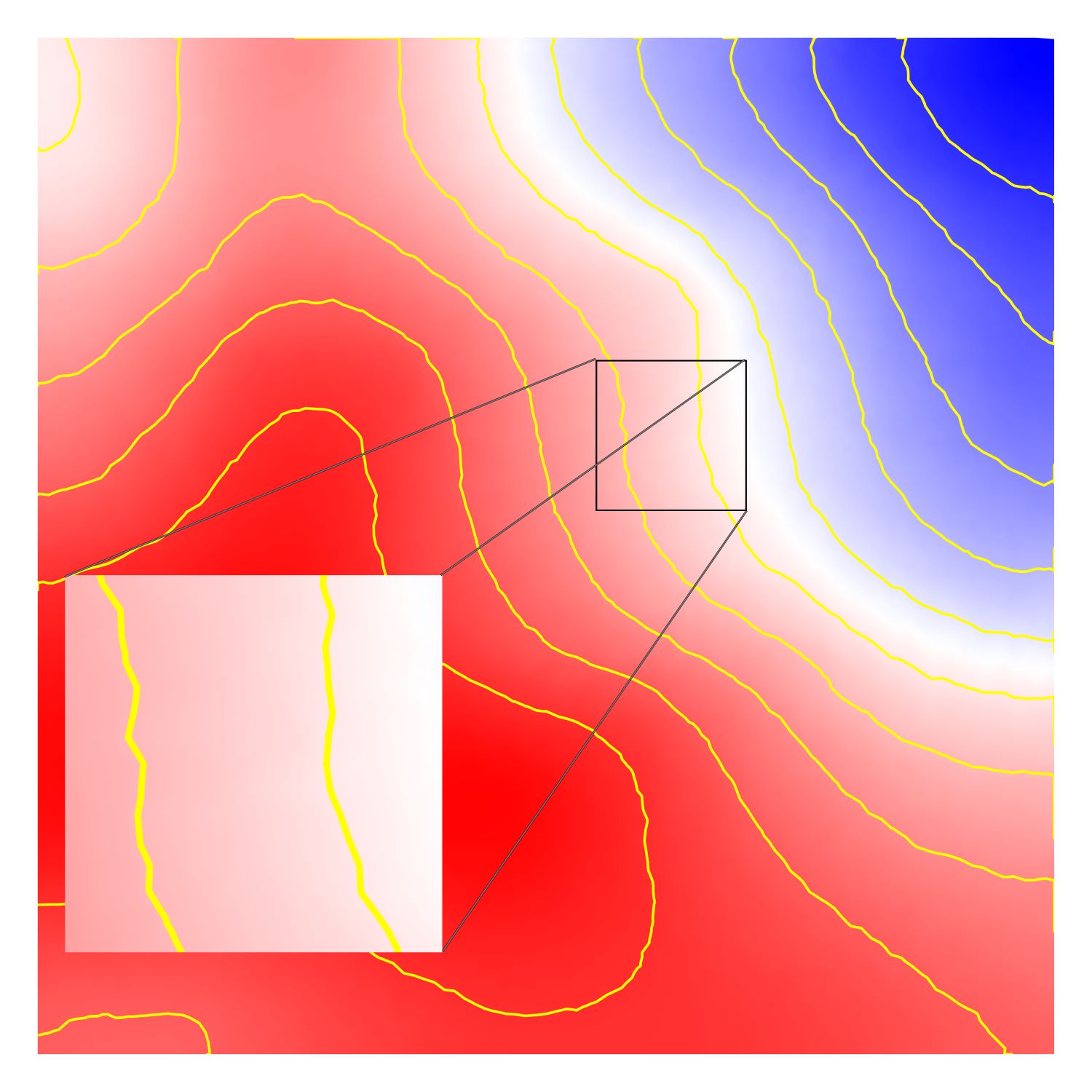}
        \subcaption{TIN smoothing w.o. angle re-weighting}\label{fig:tin_smooth_bad}
    \end{subfigure}
    \begin{subfigure}{0.49\columnwidth}
        \centering
        \includegraphics[width=\textwidth]{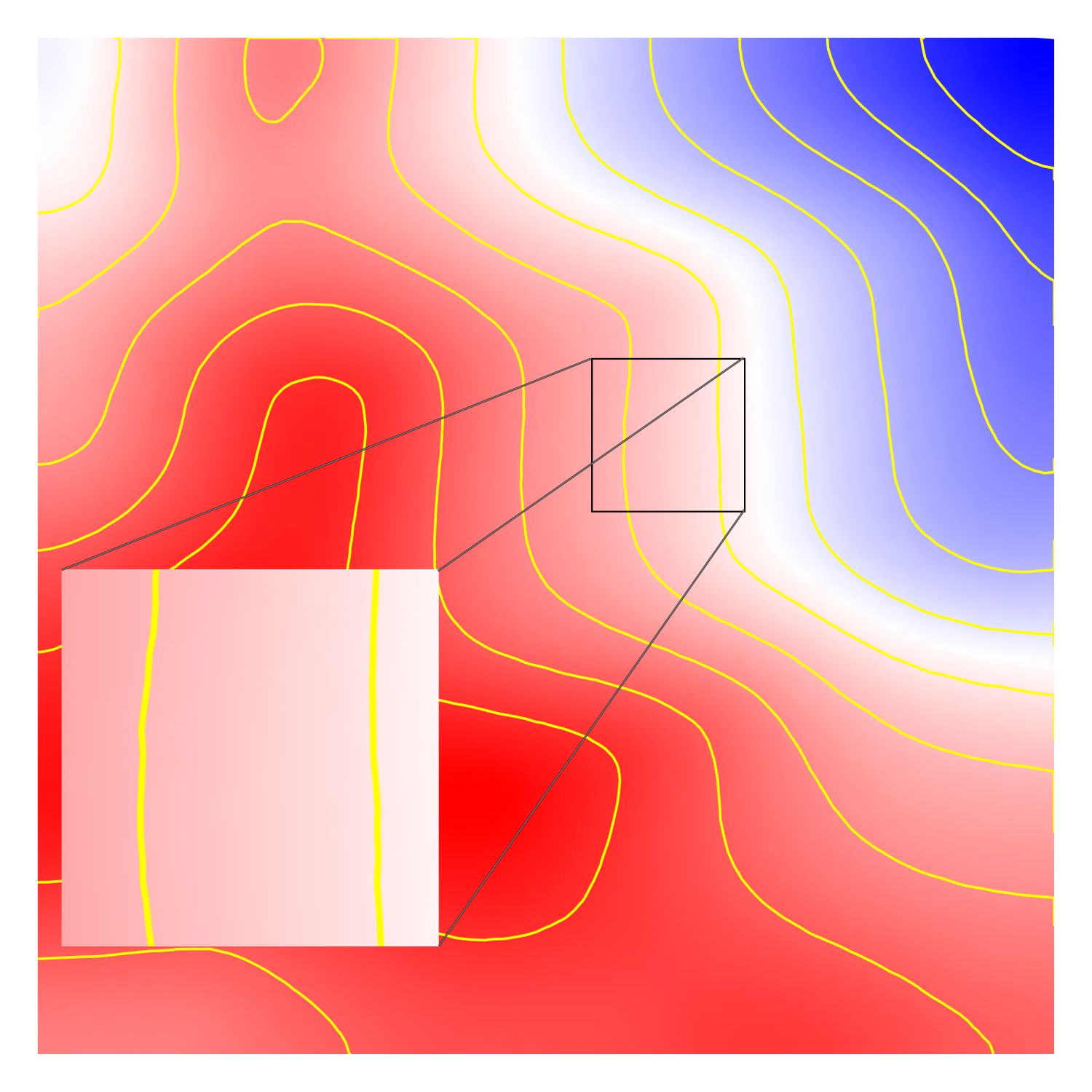}
        \subcaption{TIN smoothing}\label{fig:tin_smooth}
    \end{subfigure}
    \begin{subfigure}{0.49\columnwidth}
        \centering
        \includegraphics[width=\textwidth]{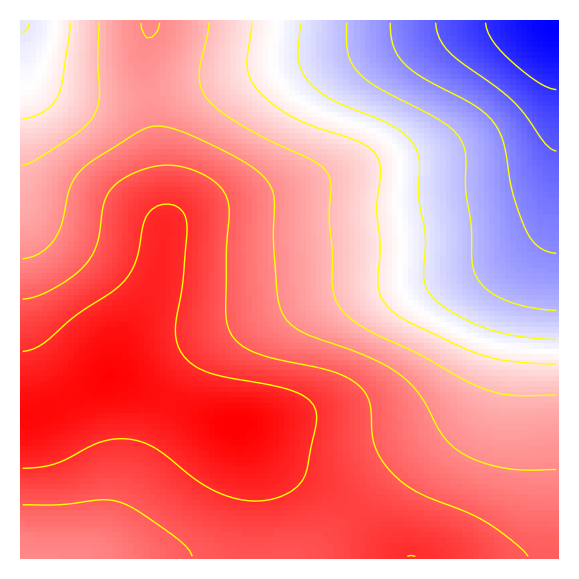}
        \subcaption{Gridded DEM smoothing}\label{fig:grid_smooth}
    \end{subfigure}
    \caption{Comparison of different smoothing results. Figure (a,b,c) progressively applies virtual neighbors and angle re-weighting techniques for the TIN smoothing. Figure (d) shows the reference image smoothing result of gridded DEM.}\label{fig:smooth_cmp}
    \Description{The figure shows the comparison of different smoothing results. The first figure shows the raw TIN smoothing result and the contour lines are rough and jitter a lot. The next (b) adds virtual neighbors and the contours become smoother, but still not as good as the benchmark in (d). Finally, (c) shows our TIN smooth result with virtual neighbors and angle re-weighting applied, and it looks closely aligned with the image smoothing result.}
\end{figure}

Differently from regular meshes, vertices adjacent to the same vertex $v$ in a TIN have different distances from $v$. The weight of an adjacent vertex diminishes significantly when it is located farther away than others, due to the application of an inverse exponential function. This phenomenon becomes more pronounced when iterations are performed with a smaller kernel size, leading to a build-up of minor errors. Figure~\ref{fig:smooth_cmp}(a) shows the TIN smoothing result when directly re-weighting adjacent vertices with an equivalent kernel size. It is noteworthy that the TIN smoothing process fails to produce a smooth surface and results in contour lines appearing disjointed rather than continuous.

\begin{figure}[t]
    \centering
    \includegraphics[width=0.7\linewidth]{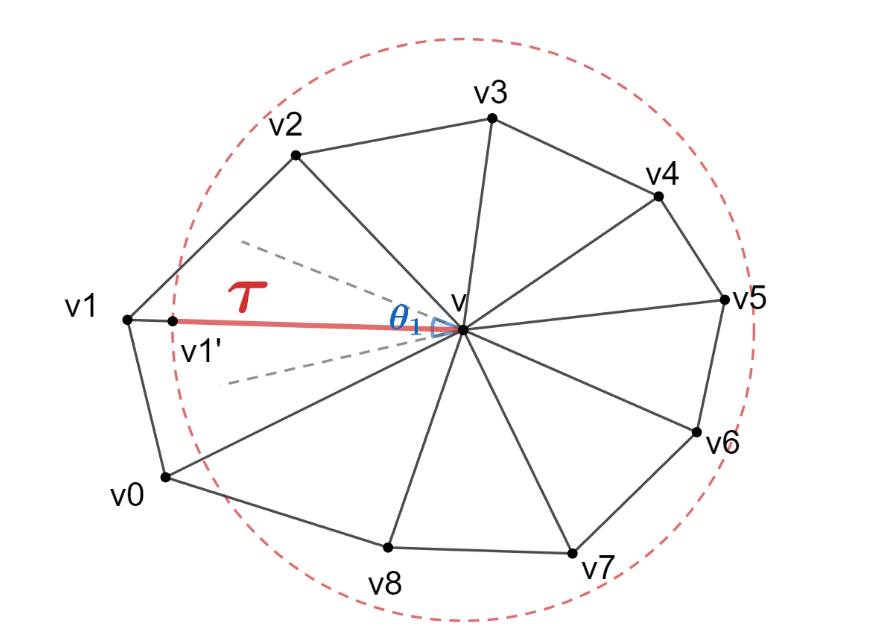}
    \caption{Virtual neighbors and angle re-weighting. For neighboring vertex $v_1$, a virtual neighbor $v_1^\prime$ is projected and $\theta_{1}$ is its angle re-weighting coefficient.}
    \label{fig:anglereweight}
    \Description{The figure shows the generation of virtual neighbors and angle re-weighting. For neighbor $v_1$, virtual neighbor $v_1^\prime$ is generated and $\theta_{1}$ is its angle re-weighting coefficient.}
\end{figure}

To rectify this, we introduce \textit{virtual neighbors} during the smoothing process. When the distance between a vertex $v$ and its adjacent vertex $v_i$ exceeds a specified threshold $\tau$, a virtual vertex is added to an edge $\overline{vv_i}$. As illustrated in Figure~\ref{fig:anglereweight}, since the distance $d(v, v_1)$ between $v$ and $v_1$ is greater than $\tau$, a virtual neighbor $v_1^\prime$ is projected onto edge $\overline{v v_1}$ at distance $\tau$ to $v$. The interpolated elevation value at this point, $z_1^\prime$, is calculated as $\tau z_1 + (1 - \tau) z$. The same principle applies to $v^\prime$ as a virtual neighbor to $v_1$. The weights of virtual neighbors can be calculated in a way similar to Equation~\ref{equa:weight}, the only difference being that the interpolated values of virtual neighbors need an extra update at the end of each smoothing iteration.

Another factor influencing the smoothing result is the spatial distribution of adjacent vertices. Unevenly distributed adjacent vertices can lead to biased weights towards regions with higher point density, and thus the smoothing result tends to be erroneous as shown in Figure~\ref{fig:smooth_cmp}(b). To address this issue, we introduce \textit{angle re-weighting}, which re-weights the weight matrix $M_{w}$ according to the angle $\theta$ covered by adjacent vertices. As shown in Figure~\ref{fig:anglereweight}, $\theta_{1}$ is calculated as the average of the angles between vertex $v_1$ and its two adjacent vertices, $v_0$ and $v_2$. Thus, the re-weighted weight matrix $M'_{w}$ is computed as:
\begin{equation}\label{equa:anglereweight}
    M'_{w}[i, j] = \frac{1}{Z^\prime} \theta_{j}\exp\left(-d(i, j)^2 / \left(2\sigma_{small}^2\right)\right)
\end{equation}
where $Z^\prime$ is the normalization factor. As demonstrated in Figure~\ref{fig:smooth_cmp}(c), incorporating the angle-based re-weighting approach improves the smoothing result, making it closely resemble the image smoothing result in Figure~\ref{fig:smooth_cmp}(d) and, consequently, the smoothing result of the real terrain.

\subsubsection{Critical points tracking and life span calculation}\label{sec:tracking}
Critical point tracking is achieved by processing the critical point transitions (equivalently, edge-flip events) between two consecutive scales in chronological order. As described in Section~\ref{sec:bg}, the type of a vertex $v_i$ is defined by its vertex signature at the timestamp $t$, denoted as $\eta_{i,t}$. For an edge $\overline{v_nv_m}$ flipping at the timestamp $t$, the critical point types at the timestamp before and after edge flipping are determined. This critical point transition can be written as:
\begin{equation}
    (\eta_{n,t-\delta}, \eta_{m,t-\delta}) \rightarrow (\eta_{n,t+\delta}, \eta_{m,t+\delta})
\end{equation}
where $\delta > 0$ represents a small change in the timestamp.

In the grid-based method by~\cite{Rocca2017}, the types of transitions of critical points are limited on regular triangle meshes with a preset edge connection. There are no $k$-fold saddles ($k>2$) since each vertex of a regular mesh has exactly six adjacent vertices. A $2$-fold saddle is treated as two simple saddles sharing the same vertex. Therefore, the transition is simplified to 32 preset cases in a transition table. However, on a TIN, the number of vertex neighbors is not constant and the existence of $k$-fold saddles ($k > 2$) makes the critical point transition much more complex. This grid-based method leaves out the complex transitions involving more than two critical points.

\begin{figure}[t]
    \centering
    \includegraphics[width=\linewidth]{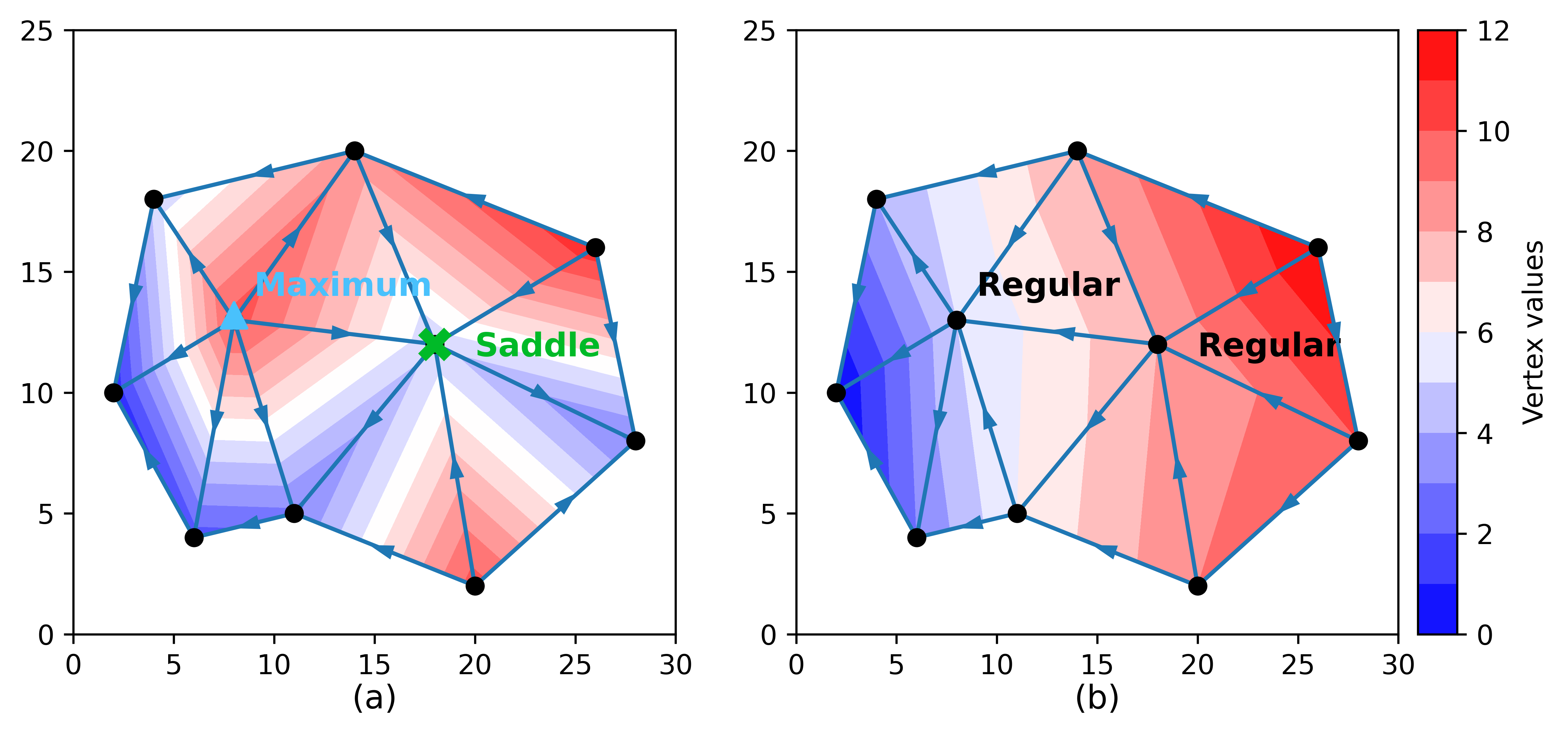}
    \caption{\textit{Maximum} and \textit{saddle} collapse to two \textit{Regular} points with the scale changes from fine (a) to rough (b). Vertex values are interpolated as face colors of triangles and arrows on edges denote the value relationship from high to low.}
    \label{fig:collapse}
    \Description{The figure shows the collapse of two critical points. Neighboring Maximum and Saddle collapse to two regular points after the scale changes from fine to rough.}
\end{figure}

In our method, we propose a generalized critical feature tracking approach that works on TINs and handles complex transitions involving $k$-fold saddles and more than two critical points.  Critical point transitions are categorized into three types: \textit{Displacement}, \textit{Appearance}, and \textit{Collapse}. For instance, the transition (\textit{maximum}, \textit{regular}) $\rightarrow$ (\textit{regular}, \textit{maximum}) denotes a displacement of a maximum through the flipping edge from the first vertex to the other vertex. For the collapse and appearance transition, by the constraint of the Poincar\'{e}-Hopf index theorem~\cite{milnor1963morse} (see Equation~\ref{equa:index}), the transition can only happen to critical points in pairs of the form \textit{(maximum-saddle)} or \textit{(minimum-saddle)}, e.g. after a \textit{Collapse} transition on a flipping edge, a \textit{maximum} collapses with a \textit{saddle} leaving two \textit{regular} points, as illustrated in Figure~\ref{fig:collapse}.

Similarly to the solution for $2$-fold saddles, a $k$-fold saddle is regarded as $k$ simple saddles that share a vertex. The transition involving $k$-fold saddles is considered separately into steps involving only a simple saddle from this vertex at a time. Each step's transition type can be determined by a matching of the critical point pair (saddle- maximum/minimum). For example, the transition \textit{(2-fold saddle, maximum)} $\rightarrow$ \textit{(regular, saddle)} involves three critical points: the collapse of the (maximum-saddle) pair and the displacement of the other saddle. To resolve the matching problem, the trace of each critical point on the TIN is considered. The ``velocity'' of each critical point is defined as its potential to move through the current flipping edge. For example, considering only the last edge that a critical point passed, if it aligns with the current flipping edge, the critical point is considered to have higher ``velocity'' to transit through this edge. Then, this critical point is matched first with the other critical point on the other extreme of the edge.

After recording all the traces of critical points, to quantitatively measure the importance of each critical point, a \textit{life span} is defined as the number of scales in which a critical point persists. We call a critical point extracted from the original TIN as an \textit{initial} critical point, and a critical point appears in other scales of the scale space as a \textit{newborn} critical point. During the evolution of the scale space, newborn critical points may collapse to the initial critical points, shortening the life span of the initial critical points.

Therefore, a \textit{life span recovery} process is necessary. A connection can be established between the initial critical points and the newborn critical points at the intermediate scales. Newborn critical points appear in pairs and are considered as the \textit{birth\_mate} of each other. Conversely, when two critical points collapse, they are considered as the \textit{death\_mate} of each other. Given an initial critical point $p$ that collapses to a newborn critical point $q$, if $q.birth\_mate$ is still alive after the collapse timestamp, $q.birth\_mate$ is considered as a substitution of $p$ and is tracked until there is another collapse.

In this way, the life span of $p$ is extended by the life span of $q.birth\_mate$. This life span recovery process can be repetitively applied: (a) until the end of the scale space is reached or (b) until the critical point (or its substitution) collapses with an initial critical point. Thus, the life span of an initial critical point $p$ is the timestamp when $p$ collapses with another initial critical point (or its substitution), or the timestamp of the last scale layer, indicating that $p$ is still alive. 

Within the proposed pipeline, we examine all the edges in the TIN across each scale layer, thus, the per-layer run time complexity is $O(E)$ where $E$ is the number of edges in the TIN. As for the memory usage complexity, it is $O(E+V)$ within Terrain trees, where $V$ stands for vertex number. According to Euler's formula~\cite{west2001introduction}, the number of edges is $O(V)$ for a connected planar graph, which is the case for a TIN. Therefore, the proposed pipeline has linear time and memory costs with respect to the number of vertices in the TIN, making it scalable for large terrain datasets.

\section{Experimental comparison}\label{sec:exp}
In this section, we evaluate the effectiveness and efficiency of our pipeline through an experimental comparison with the \textit{grid-based method}~\cite{Rocca2017}. Experiment setup and evaluation metrics are first introduced in Section~\ref{sec:expsetup}. Then, Section~\ref{sec:spotheight} shows the effectiveness of our scale-space method for TINs compared to gridded DEMs. Additionally, Section~\ref{sec:resolution} further evaluates the resolution robustness of our TIN-based method compared to the grid-based method.

\subsection{Experiment setup and evaluation metrics}\label{sec:expsetup}
Our experimental setup encompasses a comparative assessment of our extension on TINs versus prior work on gridded DEMs by~\cite{Rocca2017}. The same evaluation task, \textit{spot height selection}, is employed, in which maxima of input terrain data are extracted based on their topological prominence. The terrain datasets consist of the point cloud collected from different regions and the spot heights manually compiled by human experts to represent terrain topography. These datasets are available from the Swiss Federal Office of Topography (\href{https://www.swisstopo.admin.ch/}{swisstopo}) and the dense ground point clouds are from their sub-dataset swissSURFACE3D~\cite{swissSURFACE3D}, featuring an average of $15-20 pts/m^2$ of the terrain topological data, which are directly fed into our pipeline.

\begin{figure}[!htbp]
    \centering
    \includegraphics[width=0.9\linewidth]{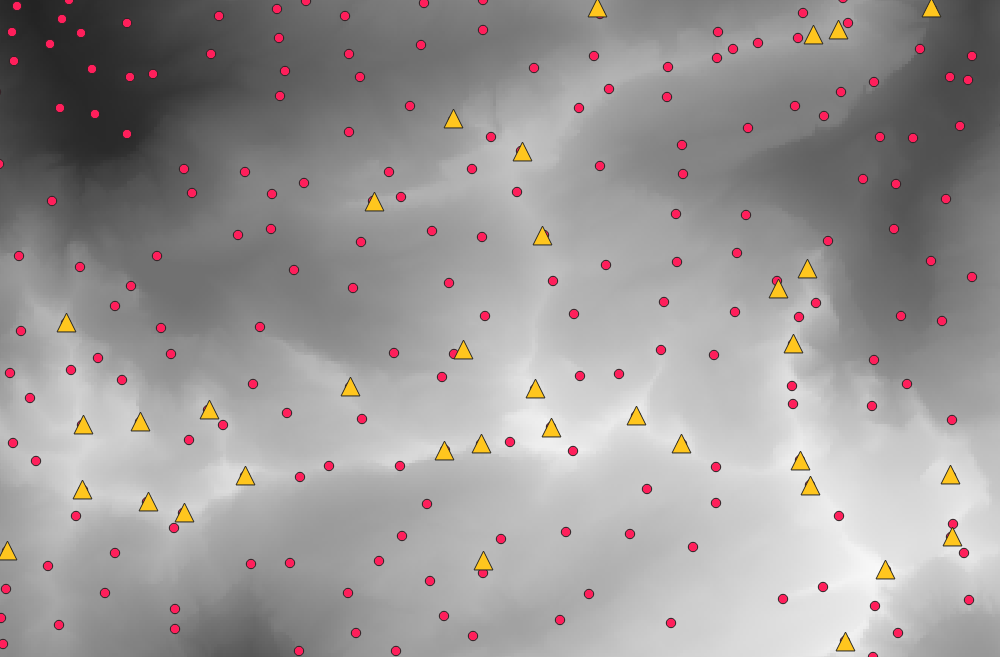}
    \caption{An example of spot heights dataset. Red circles denote spot heights and yellow triangles denote named peaks.}\label{fig:spotheights}
    \Description{The figure shows an example of the spot heights dataset. Overlying a gray-scale image of the terrain, red circles denoting spot heights and yellow triangles denoting named peaks are drawn.}
\end{figure}

Figure~\ref{fig:spotheights} is an example of the spot heights in a mountainous region. Most prominent spot heights are given names during the map generation process, denoted as \textit{named peaks}. Other spot heights are noted as \textit{summits} in the rest of the paper. Named peaks and summits together are considered the ground truth for the spot height selection task. To evaluate our method, the maxima identified by our pipeline are compared with the ground truth. As a benchmark, gridded DEMs of different resolutions are generated from the same input point clouds using the PDAL library~\cite{pdal2022}. A summary of the datasets tested in the experiment can be found in Table~\ref{table:datasets}.
\begin{table*}[ht]
    \caption{A summary of the datasets used in the experiment. Datasets are classified into two categories, mountainous and suburban, by their terrain types. \#points denotes the number of points in the raw point cloud from the swissSURFACE3D~\cite{swissSURFACE3D}.}\label{table:datasets}
    \begin{tabular}{cccccc}
        \toprule
        \textbf{Dataset} & \textbf{Dimension (km)} & \textbf{\#summits} & \textbf{\#peaks} & \textbf{\#points (million)} & \textbf{Type} \\
        \midrule
        Reichenburg      & $4 \times 4$            & 7                  & 9                & 167                         & Suburban      \\
        Sörenberg        & $8 \times 8$            & 16                 & 16               & 1529                        & Mountainous   \\
        Bannelpsee       & $12 \times 12$          & 22                 & 67               & 3548                        & Mountainous   \\
        \bottomrule
    \end{tabular}
\end{table*}

To evaluate the performance of our pipeline, the spot height selection task is formulated as a binary classification problem. Maxima are matched with peaks and summits in the ground-truth dataset. The matching distance is preset as 50 meters, which is aligned with the previous work~\cite{Rocca2017}. Matching precision is calculated as the ratio of critical maxima matched correctly (i.e., those that align with the ground truth) to the entire set of maxima. Conversely, the recall rate is the ratio of correctly selected spot heights to the total number of spot heights. The precision-recall curve (PR curve) and the $F_{\beta}$ score are used as evaluation metrics, generated by adjusting the life span filtration threshold for maxima. The $F_{\beta}$ score is a weighted harmonic mean of precision and recall rate, as per Equation~\ref{equa:fbeta}:
\begin{equation}\label{equa:fbeta}
    F_{\beta} = (1 + \beta^2) \frac{precision \times recall}{\beta^2 \times precision + recall}
\end{equation}
where $\beta$ is set to 0.5 to prioritize precision twice as important as the recall rate. In addition, an alternative evaluation metric $dist_{avg}$ is introduced, which denotes the average distance from the critical points to the corresponding ground-truth spot heights. It measures the precision and robustness of the critical point identification.

To evaluate the computation efficiency of our pipeline, we measure the runtime and peak memory usage of the critical point tracking stage. Point cloud preprocessing, TIN generation, and TIN smoothing are excluded from the evaluation as they depend on various factors such as the scale of the input point cloud, the targeted resolution, and the topographical complexity of the terrain. Therefore, these stages are processed offline for each dataset. The smoothed elevation value of each vertex is stored as pre-computed vertex properties. Experiments are conducted on a workstation with an Intel Core i9-12900K CPU, 64GB of RAM, and an NVIDIA GeForce RTX 3080Ti GPU.

\subsection{Spot height selection results}\label{sec:spotheight}
For the input dense point cloud data, we generate both gridded DEMs and TINs of a similar number of vertices by tuning the FPS sampling rate. From the results summarized in Table~\ref{table:result}, our pipeline's runtime and peak memory usage are close to those of the grid-based method since our extension on TINs does not introduce additional computational overhead. In line with our analysis, the time elapse of the scale-space critical point tracking stage reveals a linear correlation of the runtime and peak memory usage with the total number of edges in a TIN. The random memory access time for TINs is slightly longer than that for gridded DEMs.

\begin{table*}[ht]
    \caption{Evaluation results of the grid-based and TIN-based methods on the datasets. For both methods, columns represent different metrics: number of edges $\#edges$, average critical point matching distance $dist_{avg}$, and best $F_{\beta}\ score$. The time elapse $t_{ss}$ and peak memory usage $Mem_{peak}$ are recorded for the critical point tracking stage.}\label{table:result}
    \begin{tabular}{c|c|c|c|c|c|c|c|c|c|c|c}
        \hline
         Datasets           & \multicolumn{1}{c|}{Grid}          & \multicolumn{2}{c|}{\#edges} & \multicolumn{2}{c|}{$dist_{avg}$ (m)} & \multicolumn{2}{c|}{$F_{\beta}\ score$} & \multicolumn{2}{c|}{$t_{ss}$ (s)} & \multicolumn{2}{c}{$Mem_{peak}$ (MBs)}                                                             \\
        \cline{3-12}
                    & \multicolumn{1}{c|}{cell size (m)} & Grid                         & TIN                                   & Grid                                    & TIN                               & \multicolumn{1}{l|}{Grid}              & \multicolumn{1}{l|}{TIN} & Grid   & TIN   & Grid  & TIN   \\
        \hline
        Reichenburg & 16                                 & 186,501                      & 170,265                               & 18.20                                   & \textbf{4.85}                     & 0.79                                   & \textbf{0.92}            & 39.33  & 36.52 & 24.38 & 22.75 \\
        Sörenberg   & 40                                 & 119,201                      & 112,000                               & 30.38                                   & \textbf{12.79}                    & 0.85                                   & \textbf{0.87}            & 26.077 & 27.81 & 17.35 & 17.00 \\
        Bannelpsee  & 40                                 & 268,801                      & 255,945                               & 32.18                                   & \textbf{14.19}                    & 0.77                                   & \textbf{0.79}            & 56.88  & 63.43 & 32.65 & 32.04 \\
        \hline
    \end{tabular}
\end{table*}

\begin{figure}[!htbp]
    \centering
    \begin{subfigure}{0.49\columnwidth}
        \centering
        \includegraphics[width=0.95\linewidth]{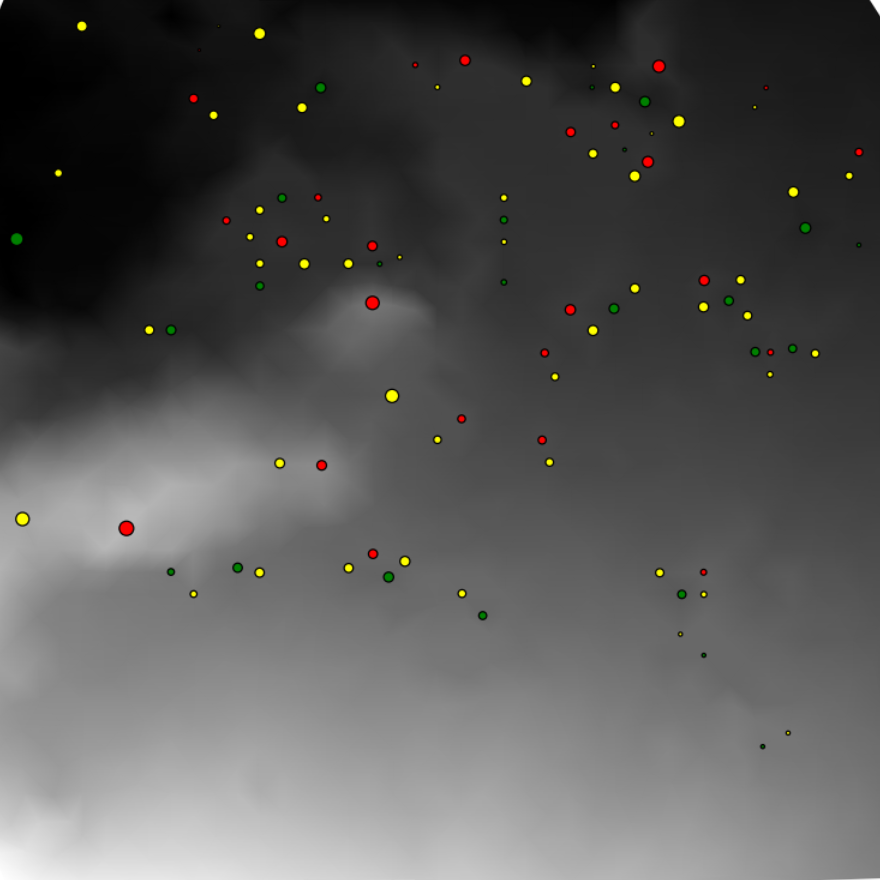}
        \subcaption{TIN - $1,126$ vertices}
    \end{subfigure}
    \begin{subfigure}{0.49\columnwidth}
        \centering
        \includegraphics[width=0.95\linewidth]{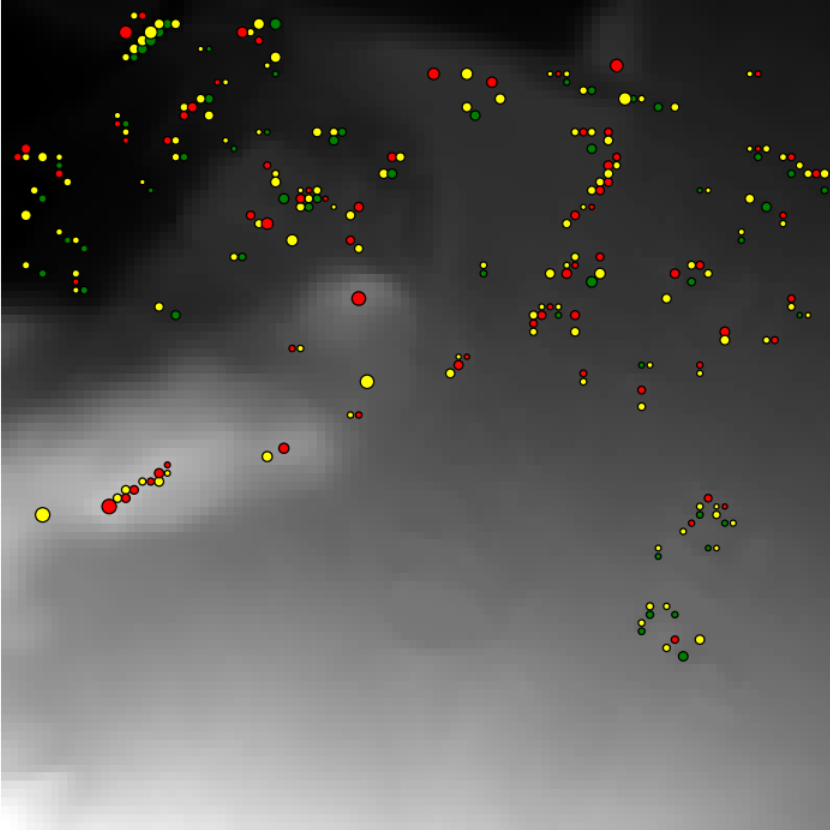}
        \subcaption{Grid - $6,400$ vertices}
    \end{subfigure}
    \begin{subfigure}{0.49\columnwidth}
        \centering
        \includegraphics[width=0.95\linewidth]{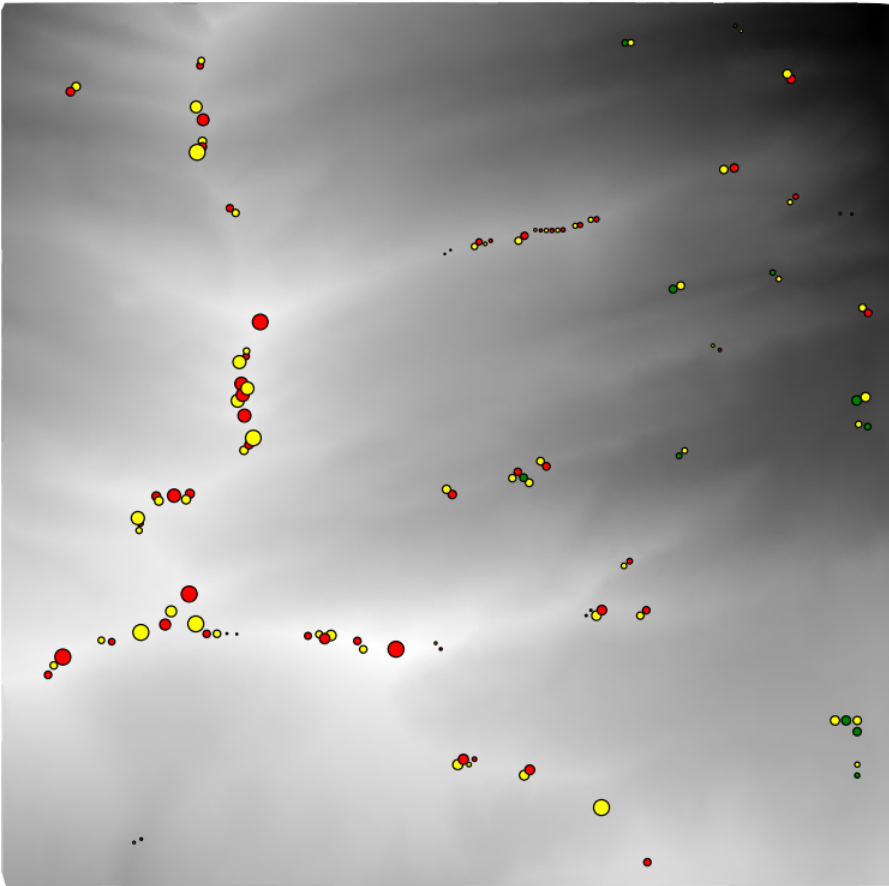}
        \subcaption{TIN - $11,447$ vertices}
    \end{subfigure}
    \begin{subfigure}{0.49\columnwidth}
        \centering
        \includegraphics[width=0.95\linewidth]{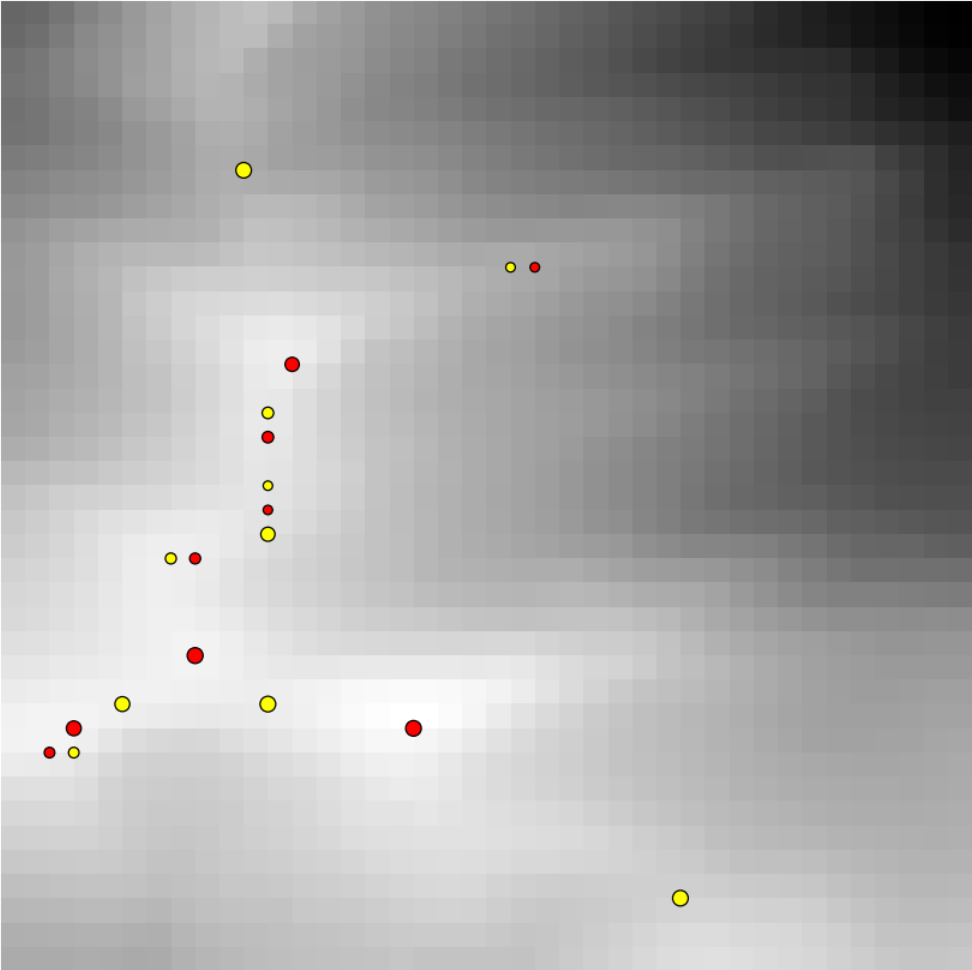}
        \subcaption{Grid - $1,600$ vertices}
    \end{subfigure}
    \caption{Visual comparison of two zoomed-in sub-regions of the Reichenburg dataset. Among them, (a,b) are of an urban region with fewer terrain changes. (c,d) are of a mountainous region with complex terrain features. Maxima, minima, and saddles are marked in red, green, and yellow circles with sizes proportional to the life span of the critical points.}
    \label{fig:urban_mount}
    \Description{Visual comparison of two zoomed-in sub-regions of the Reichenburg dataset. Maxima, minima, and saddles are marked in red, green, and yellow circles with sizes proportionally to the life span of the critical points. In suburban regions, TIN (a) has less and meaningful critical points than gridded DEM (b). In the mountainous region, TIN (c) recognizes much more critical points more accurately and depicts the terrain details better than gridded DEM (d).}
\end{figure}

The preprocessing stage plays an important role in our pipeline in taking advantage of TINs. Taking the $4 \times 4\ km^2$ suburban region Reichenburg dataset as an example, the number of vertices and edges of the TIN are $57,065$ and $170,265$ respectively, while the gridded DEM has $62,500$ vertices and $186,501$ edges. With the help of our Patch-based extension of the FPS (PFPS) method (\textit{the sidelength of patches is set empirically as 20 meters}), the variable-resolution capability of the TIN allows the allocation of more vertices in the mountainous area than in the flat one. A small adjustment is applied in the grid-based method to flat edges as a tie-breaking solution~\cite{Rocca2017}. This solution, however, introduces artificial critical points within the flat regions, called \textit{artifacts}, to the model as seen in Figure~\ref{fig:urban_mount}(b). While our method adopts the same tie-breaking method in flat areas, the number of vertices retained within flat regions is significantly reduced by our PFPS procedure, thereby reducing the occurrence of artifacts. Furthermore, the frequency of edge flip events within flat areas decreases due to the reduced number of edges subjected to the smoothing process.

For a more detailed comparison, Figure~\ref{fig:urban_mount} provides a zoomed-in view of the sub-regions. Within the same dataset, different regions are of the same resolution in the gridded DEM. In the urban region (Figures~\ref{fig:urban_mount} (a-b)), more artifacts are introduced in the gridded DEM, because of the high-resolution grid and small differences between neighboring vertices. The number of critical points reduces from $226$ (in the gridded DEM) to $98$ (in the TIN), while the same prominent critical points are recognized. However, in the mountainous sub-region shown in Figures~\ref{fig:urban_mount} (c-d), the gridded DEM provides much fewer surface details with only $1,600$ vertices, and fewer critical points are identified from the gridded DEM compared to the TIN, which contains $11,447$ vertices in the same region. The gridded DEM, constrained by its fixed grid size, fails to adapt to the varying distribution of topographical features, thereby leading to inefficient utilization of computational resources. On the other hand, the variable-resolution capability of TIN enables more accurate locating and tracking of critical points in scale space.

\begin{figure}[!htbp]
    \centering
    \begin{subfigure}{\columnwidth}
        \centering
        \includegraphics[width=0.95\linewidth]{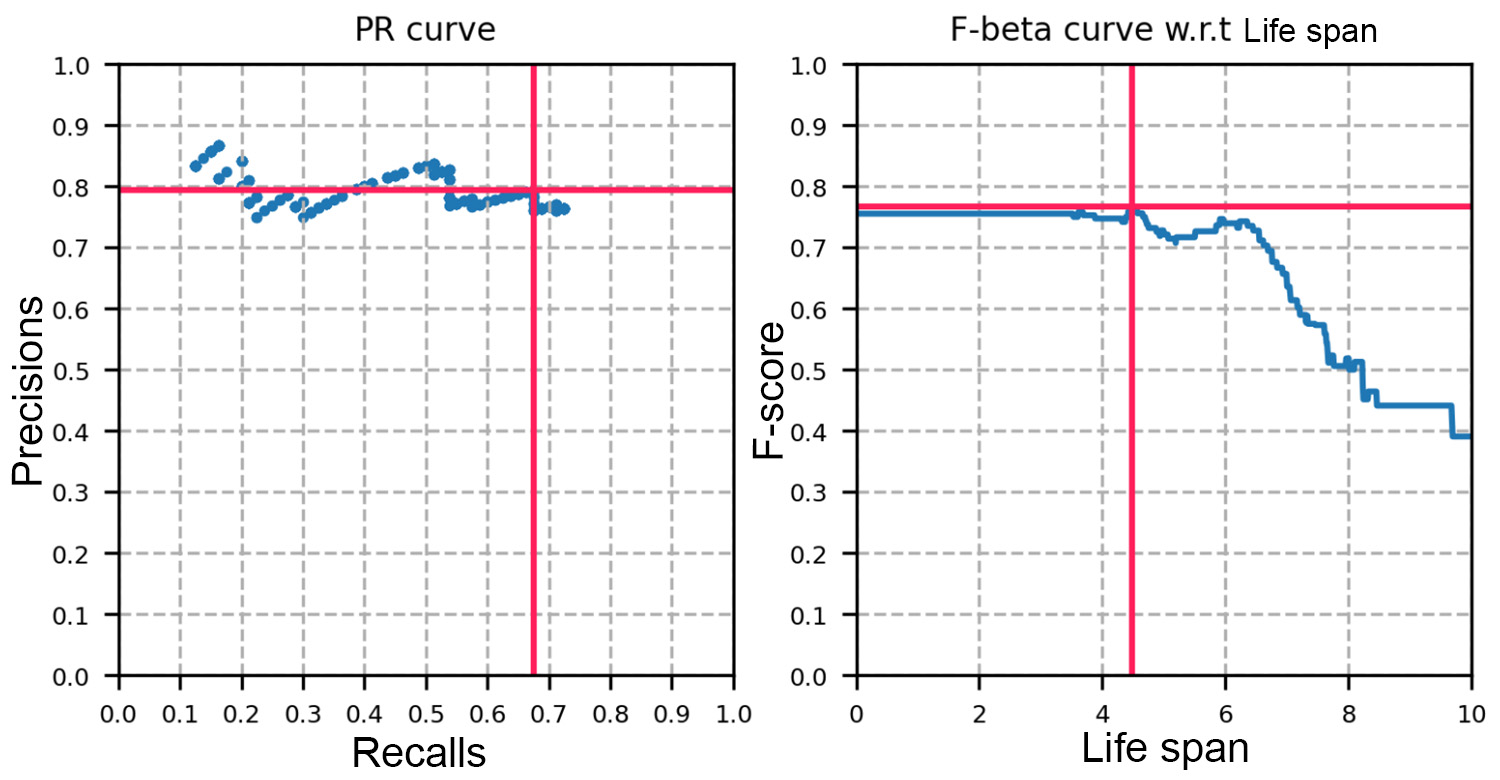}
        \subcaption{Gridded DEM - best $F_{\beta} = 0.77$}
    \end{subfigure}
    \begin{subfigure}{\columnwidth}
        \centering
        \includegraphics[width=0.95\linewidth]{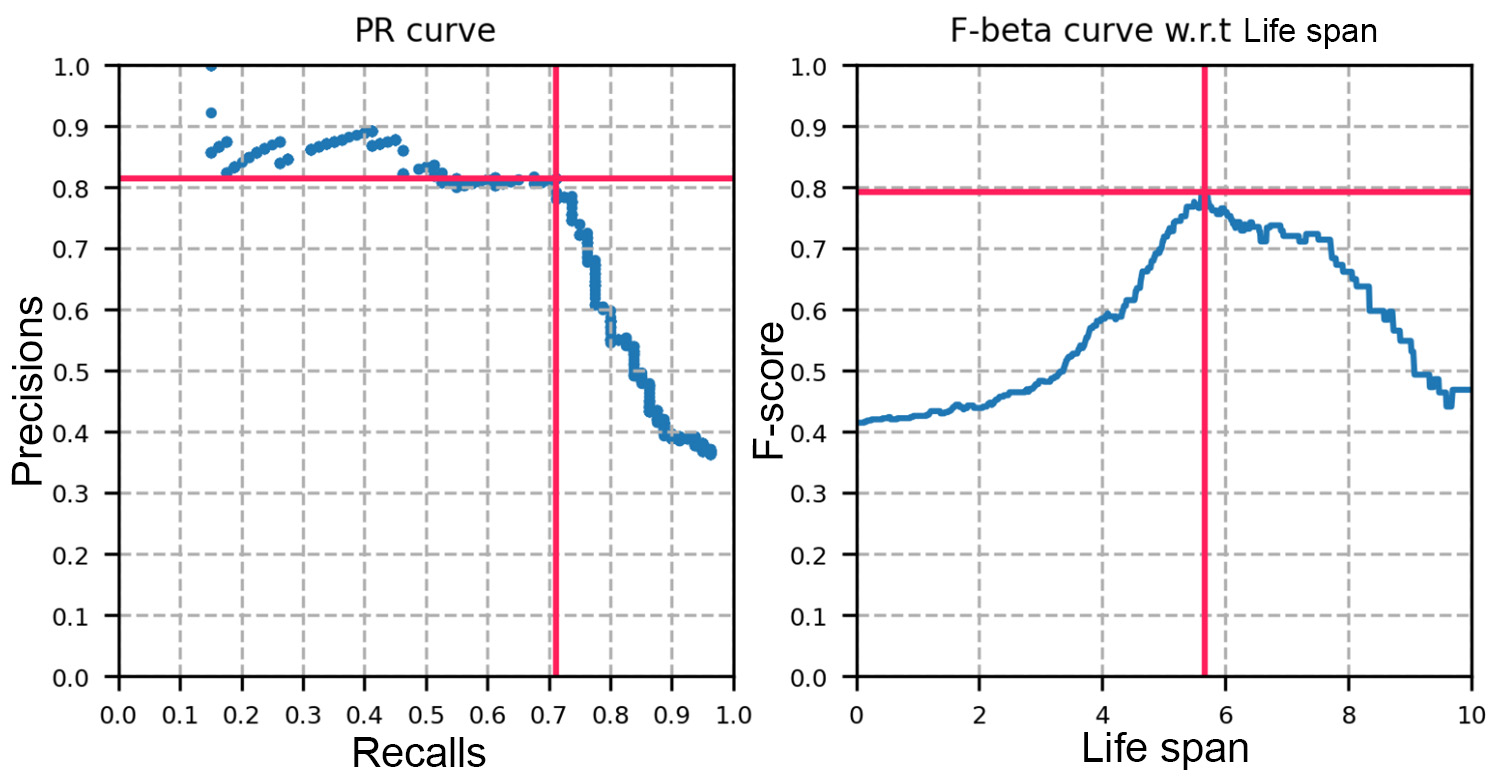}
        \subcaption{TIN - best $F_{\beta} = 0.79$}
    \end{subfigure}
    \caption{PR curve and F-score graph of the gridded DEM and TIN methods on the mountainous regions Bannelpsee.}\label{fig:prf}
    \Description{PR curve and F-score graph of the Gridded DEM and TIN methods on the mountainous regions Bannelpsee. TIN's PR curve manages to reach the higher recall rate region while gridded DEM's fails.}
\end{figure}

For a quantitative evaluation, the PR curve and $F_{\beta}$ score graph of the grid-based and TIN-based methods are drawn respectively. For example, Figure~\ref{fig:prf} illustrates the result derived from a mountainous region near Bannelpsee. It is worth noting that the gridded DEM fails to identify some of the critical maxima as the recall rate stops at $0.73$. In contrast, the TIN-based method's PR curve covers higher recall rate regions and the best $F_{\beta}$ score is $0.79$ compared to $0.77$ of the grid-based method. The high resolution of the TIN in regions with more terrain features facilitates the recognition of more critical points in the input data and, thus, more spot heights.

As shown in Table~\ref{table:result}, our TIN-based method exhibits a slight improvement in the $F_{\beta}$ score in mountainous regions, with increases of $2.35\%$ in Sörenberg and $2.60\%$ in Bannelpsee, and a large improvement in a suburban region, with an increase of $16.5\%$ in Reichenburg, compared to the grid-based method when the numbers of edges are similar in two terrain models. This highlights the advantages of the TIN-based method, as it can accurately identify critical points and minimize the risk of artifacts. Furthermore, leveraging the benefits of the PFPS method and flexible resolution, the average distance $dist_{avg}$ of the TIN-based method is considerably smaller than that of the grid-based method. For instance, in the mountainous region Bannelpsee, $dist_{avg}$ of the TIN-based method is $14.19 m$, while that of the grid-based method is $32.18 m$. In the suburban region Reichenburg, the $dist_{avg}$ of the TIN-based method is $3.75$ times smaller than that of the grid-based method.

\subsection{Resolution robustness evaluation}\label{sec:resolution}
We evaluate the robustness of our TIN-based method compared to the grid-based method under varying resolutions. For clarity, we define resolution as the side length of the grid cell in the case of the gridded DEM. To make a fair comparison, for the input point dataset, we generate TIN with a similar number of vertices to the gridded DEM. This allows us to compare the resolution robustness under a similar experimental configuration. Without loss of generality, we use the Bannelpsee dataset as an example to show the comparisons between these two terrain models. Experiments on other datasets show similar results and are not listed here.

\begin{figure}[t]
    \centering
    \includegraphics[width=\linewidth]{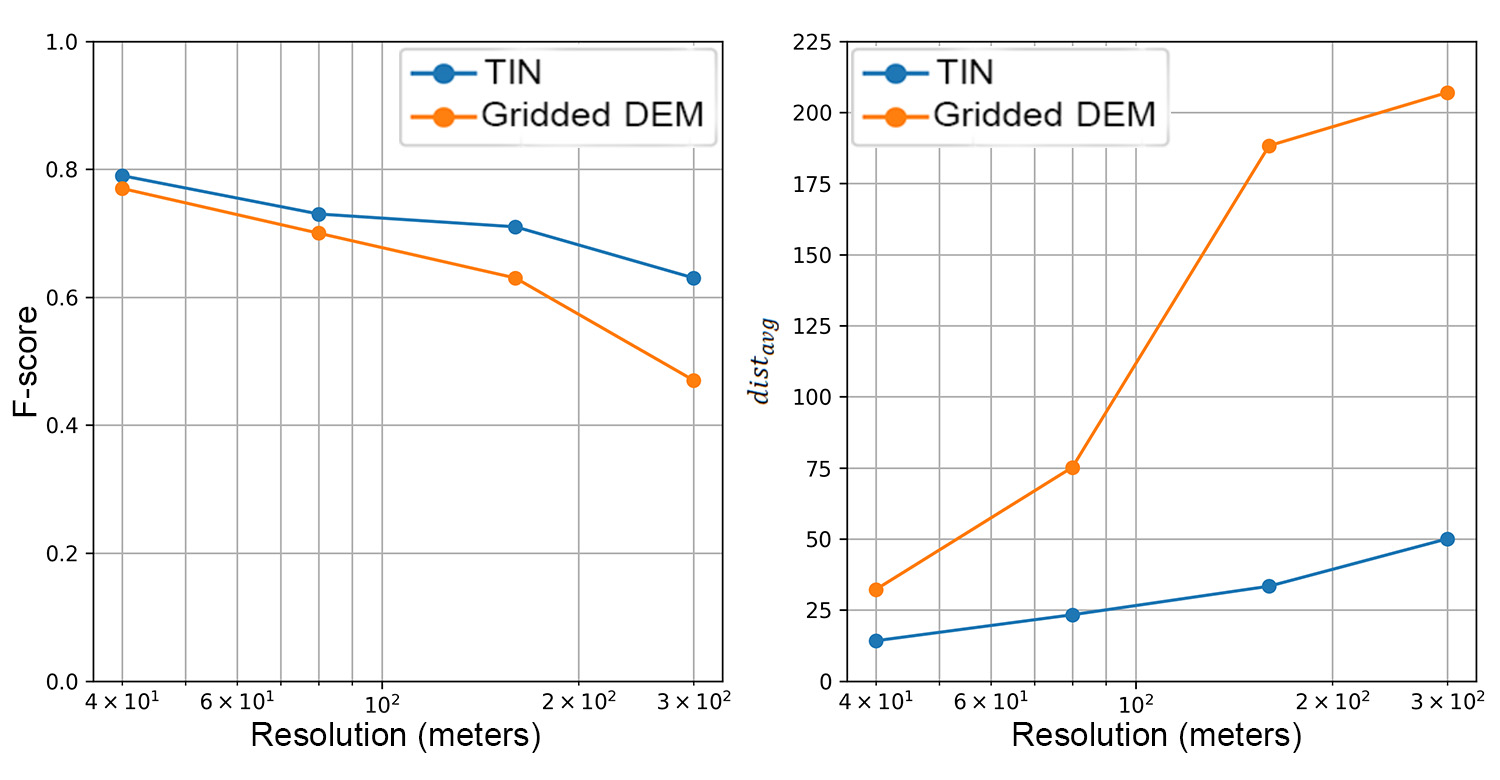}
    \caption{Resolution robustness comparison between grid-based and TIN-based methods in Bannelpsee. $F_{\beta}\ score$ (left) and $dist_{avg}$ (right) are plotted as lines. $x$-axes are in log scale.}\label{fig:resolution}
    \Description{Resolution robustness comparison between grid-based and TIN-based methods in Bannelpsee. $F_{\beta}\ score$ (left) and $dist_{avg}$ (right) are plotted as lines. $x$-axes are in log scale. TIN has better results than gridded DEM and this advance further increases when the resolution of the grid further increases.}
\end{figure}

In Figure~\ref{fig:resolution}, the $x$-axis shows the resolution changing from $40m$ to $320m$ and the $y$-axis represents the $F_{\beta}$ score and $dist_{avg}$. The graphs show that our TIN-based pipeline has superior resolution robustness compared to the grid-based method, especially when there is limited memory and the grid cell size is large. Our curvature-based sampling strategy ensures that regions with sparser topographical information are subjected to greater downsampling. Since critical points do not appear on a flat surface, fewer vertices in the flat areas will not affect the identification of critical points in the TIN. In contrast, gridded DEMs' resolution directly impacts the initial identification of critical points, and consequently, the final critical point tracking results. Besides, $dist_{avg}$ also grows inversely proportional as the resolution diminishes. These observations show that, when a suitable down-sampling strategy is applied, TINs have better resolution robustness compared to gridded DEMs. Visualizations of gridded DEMs and TINs at different resolutions are provided in Appendix~\ref{appendix:resolution} (Figure~\ref{fig:tin_grid_res}) for further comparison.

\section{Discussion}\label{sec:discuss}
We have introduced a new pipeline for scale-space analysis on TINs to identify and track critical points, exploiting the advantages of TINs over gridded DEMs. TINs are better alternatives to gridded DEMs when the input point data are distributed irregularly, due to their inherent variable-resolution representation capability. Generated from raw point cloud data, a TIN provides a natural representation of the terrain surface, obviating the need for interpolation as required to convert a point cloud into a gridded DEM.

The variable-resolution property of TINs avoids redundant points in flat areas, which contributes to the scale-space pipeline discussed in this paper. Compared to gridded DEMs, TINs have fewer vertices in flat areas, and thus, fewer artificial critical points in such areas, as discussed in Section~\ref{sec:spotheight}. Besides, this property also helps to reduce the number of edge flip events that need to be tracked, thereby reducing both the computational time and memory usage in the scale-space analysis stage.

\begin{figure}[t]
    \centering
    \includegraphics[width=\linewidth]{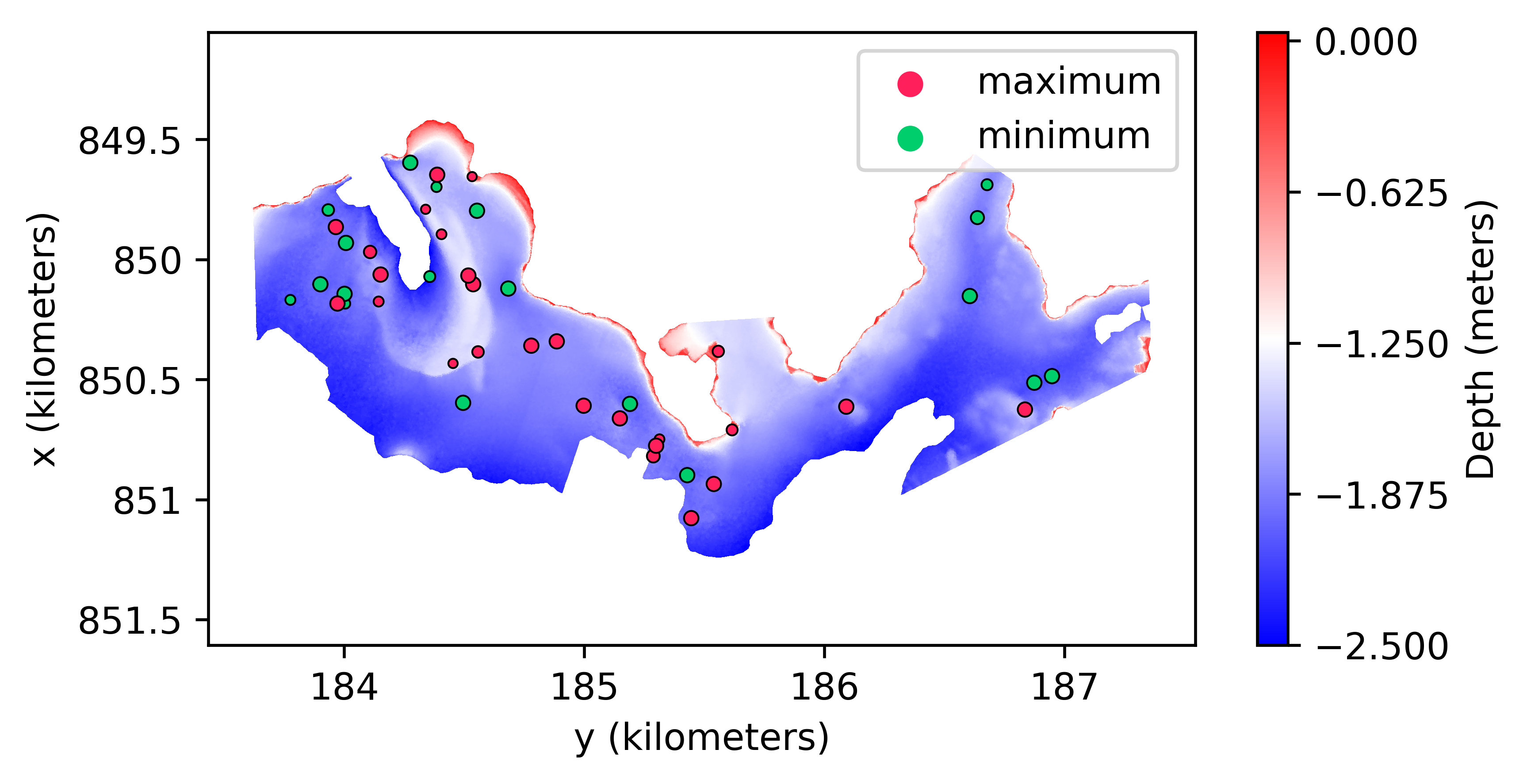}
    \caption{Maxima and minima from the bathymetry dataset of Swanquarter Bay with life span values larger than 6. Circle size represents the life span value of each critical point and can be used as a metric of importance.}\label{fig:bathymetric}
    \Description{Maxima and minima from the bathymetry dataset of Swanquarter Bay with life span values larger than 6. Circle size represents the life span value of each critical point and can be used as a metric of importance. Maxima and minima are well separated and aligned with the bathymetry surface details.}
\end{figure}

Moreover, TINs are more suitable for datasets with irregular boundaries, like coastlines in bathymetry data.  Figure~\ref{fig:bathymetric} shows one example of applying our TIN-based pipeline to a bathymetry Light Detection and Ranging (LiDAR) dataset from the National Oceanic and Atmospheric Administration (NOAA)~\cite{NOAA2024coastal}. This dataset describes the offshore bathymetric bottom surface in the vicinity of Swanquarter Bay. The irregular coastline and ubiquitous occurrence of regions without data points make it difficult for a grid-based method to be applied, as there are many cells without a value. These cells not only make steps in the scale-space method, such as smoothing and tracking, challenging or even unfeasible, but also add to the computational complexity.

For a locally evenly distributed TIN, we demonstrated that Gaussian smoothing can be effectively approximated by considering only the adjacent vertices of each vertex. Techniques such as virtual neighbors and angle-based re-weighting methods, pivotal to the smoothing process, can improve the smoothing results of variable-resolution TINs, as discussed in Section~\ref{sec:smoothing}.

Taking advantage of a TIN, the terrain surface can be modeled by an \textit{adaptive-resolution TIN}, in which the regional density of vertices is proportional to the estimated curvature of the terrain surface. This adaptive-resolution TIN can be generated by the curvature-based down-sampling method and it preserves more geometric and topological information. Furthermore, when down-sampled a lot to a sparse point set, TINs prove to be more resilient to resolution changes as discussed in Section~\ref{sec:resolution}. This property is particularly critical in cases where the input dataset is very large and computational resources are limited.

In future developments of this work, we plan to explore other TIN smoothing options to see if they can significantly improve the extraction of critical features. In the current pipeline, we parallelize the computation of the TIN smoothing operation which evidently improves computing performance. As a future work, we aim to parallelize more steps of the pipeline to parallel computing. Moreover, although in this paper we focused on applying our pipeline to the selection of spot heights, this pipeline can be applied to other fields where topological features of a scalar field are of interest. In our next step, we plan to explore additional opportunities to apply this pipeline. 

\begin{acks}
This work has been supported by the US National Science Foundation
under grant numbers IIS-1910766, IIS- 20-41415, and IIS-21-14451.
\end{acks}

\newpage
\bibliographystyle{ACM-Reference-Format}
\bibliography{ref}

\newpage
\appendix
\section{Comparison between different down-sampling methods}\label{appendix:downsampling}
\begin{figure}[!htbp]
    \centering
    \begin{subfigure}{0.49\columnwidth}
        \centering
        \includegraphics[width=\textwidth]{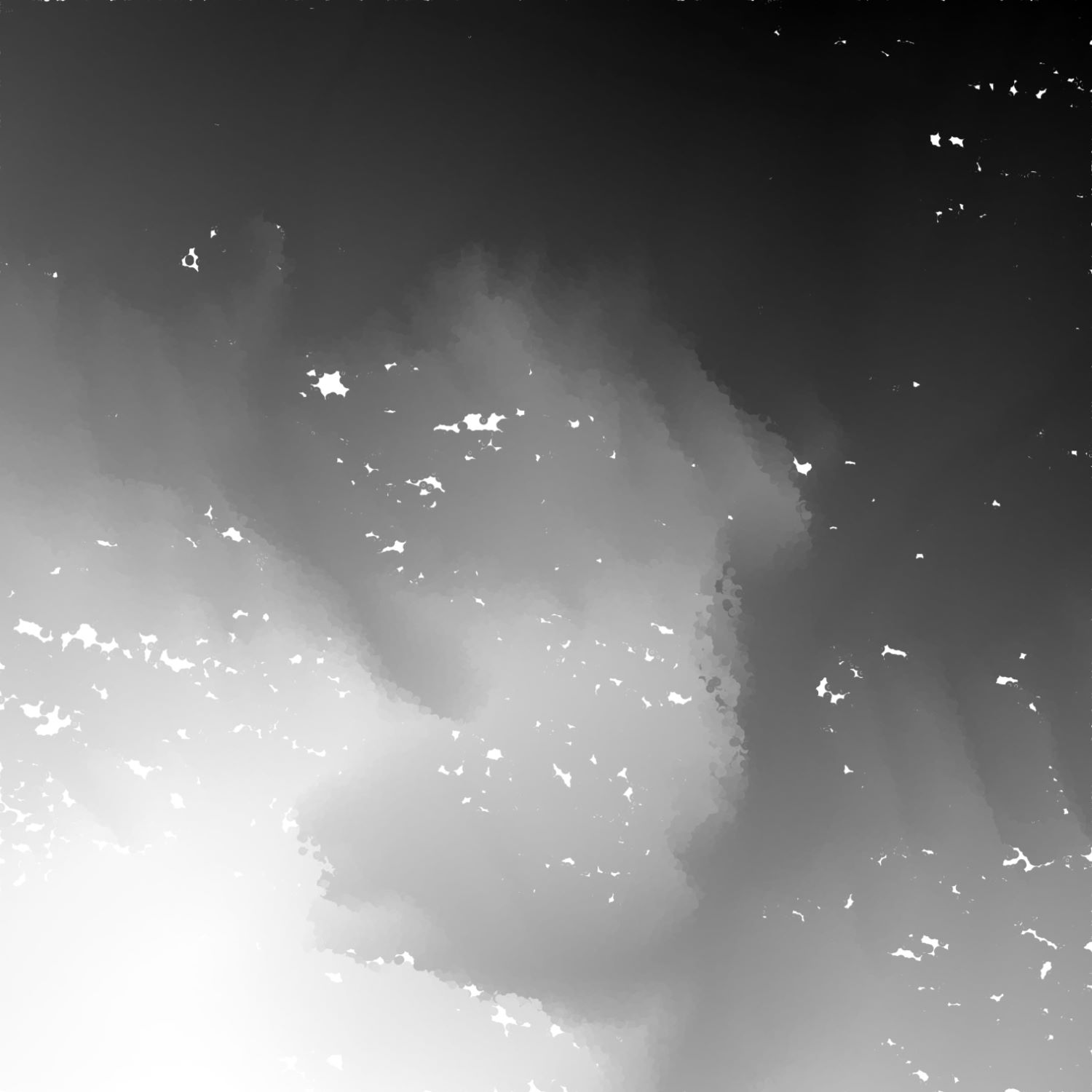}
        \subcaption{Ground points}
    \end{subfigure}
    \hfill
    \begin{subfigure}{0.49\columnwidth}
        \centering
        \includegraphics[width=\textwidth]{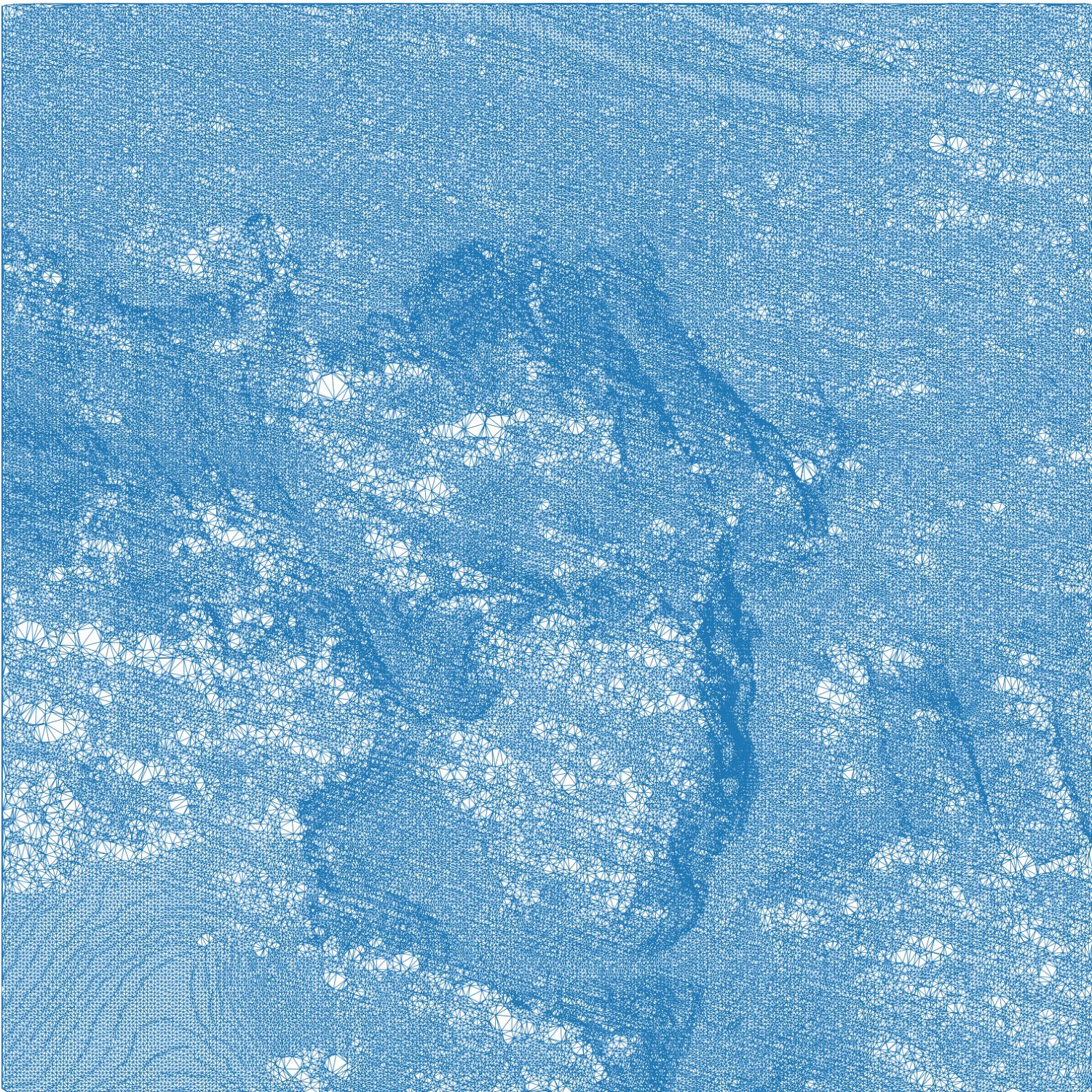}
        \subcaption{TIN - Without down-sampling}
    \end{subfigure}
    \begin{subfigure}{0.49\columnwidth}
        \centering
        \includegraphics[width=\textwidth]{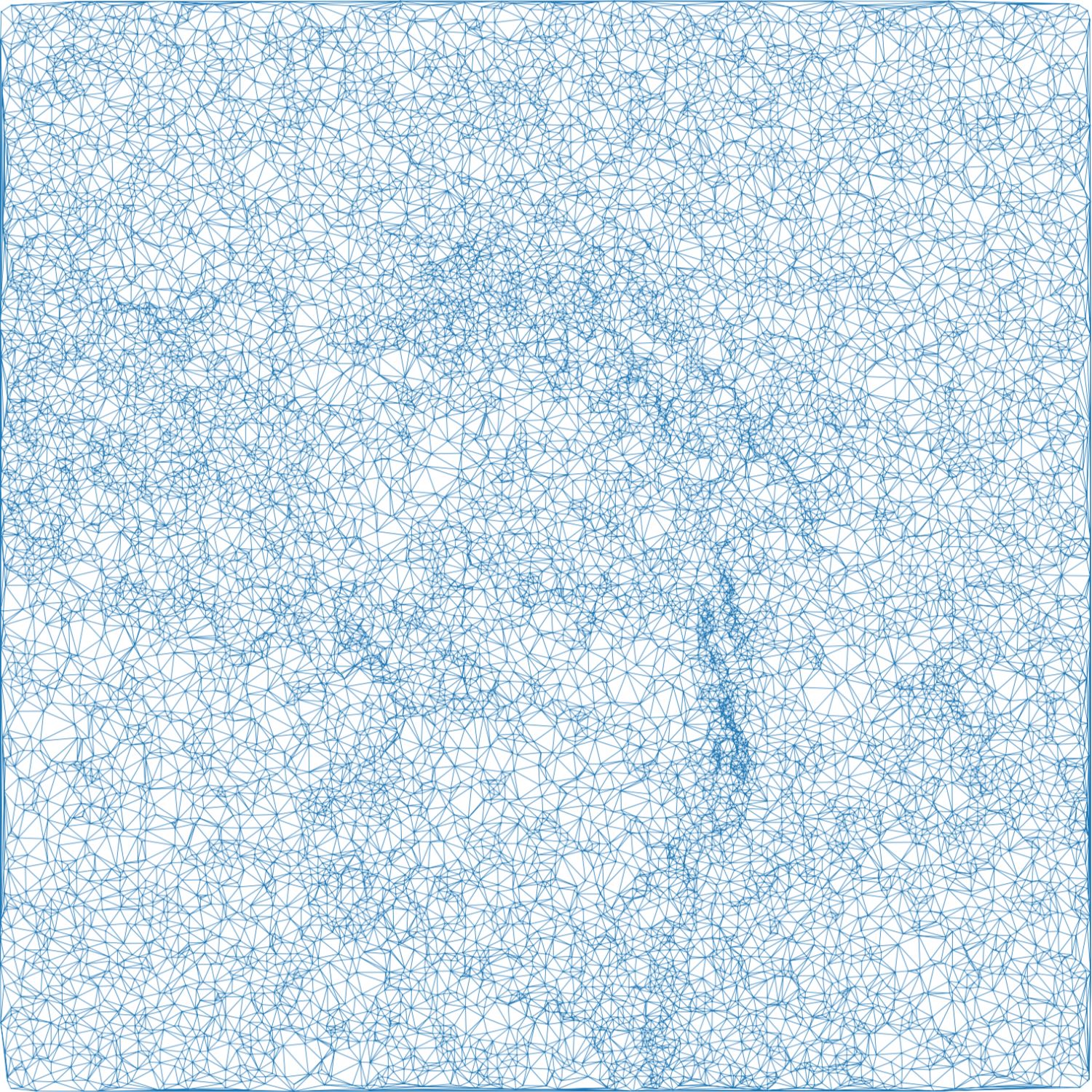}
        \subcaption{TIN - Random down-sampling}
    \end{subfigure}
    \hfill
    \begin{subfigure}{0.49\columnwidth}
        \centering
        \includegraphics[width=\textwidth]{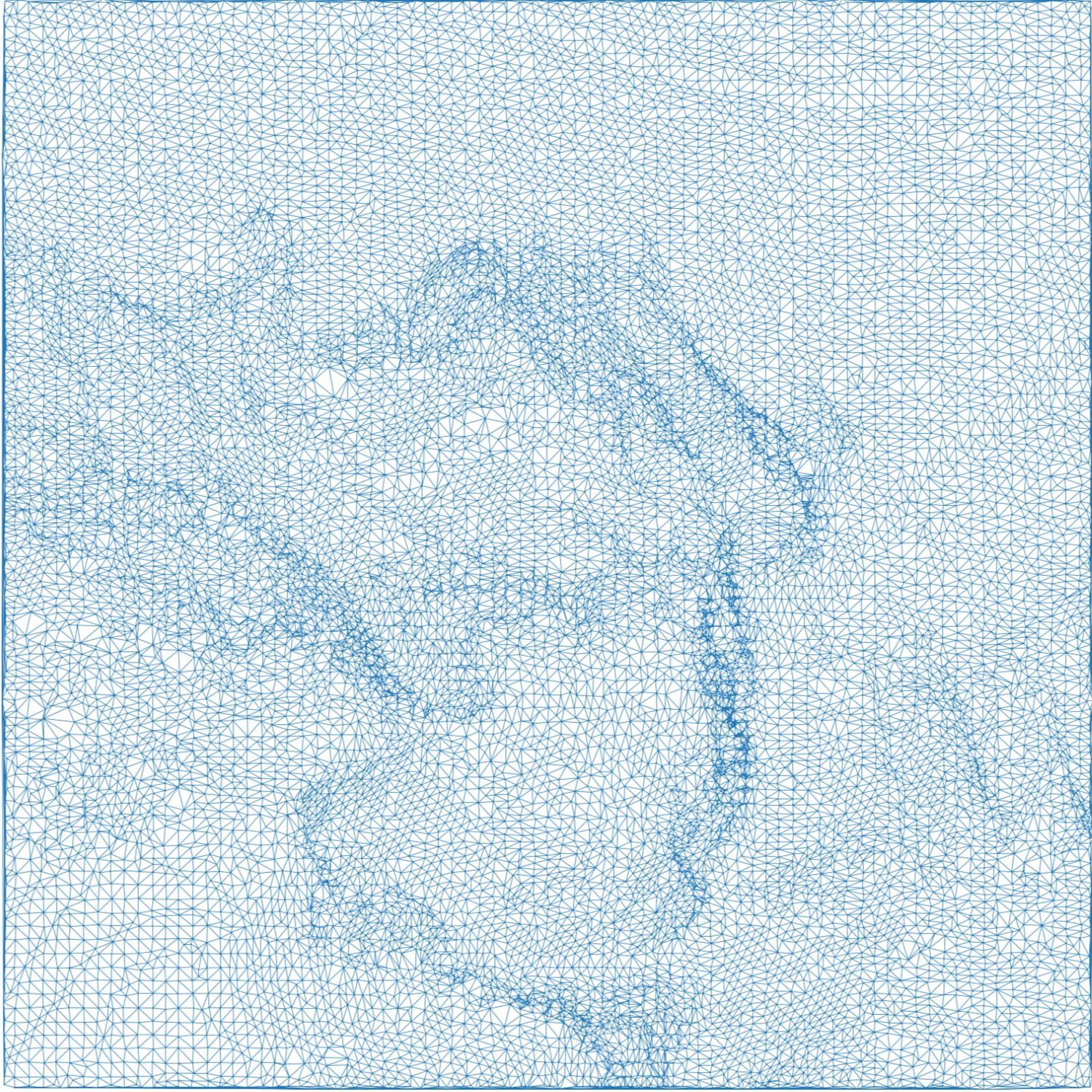}
        \subcaption{TIN - Voxel down-sampling}
    \end{subfigure}
    \begin{subfigure}{0.49\columnwidth}
        \centering
        \includegraphics[width=\textwidth]{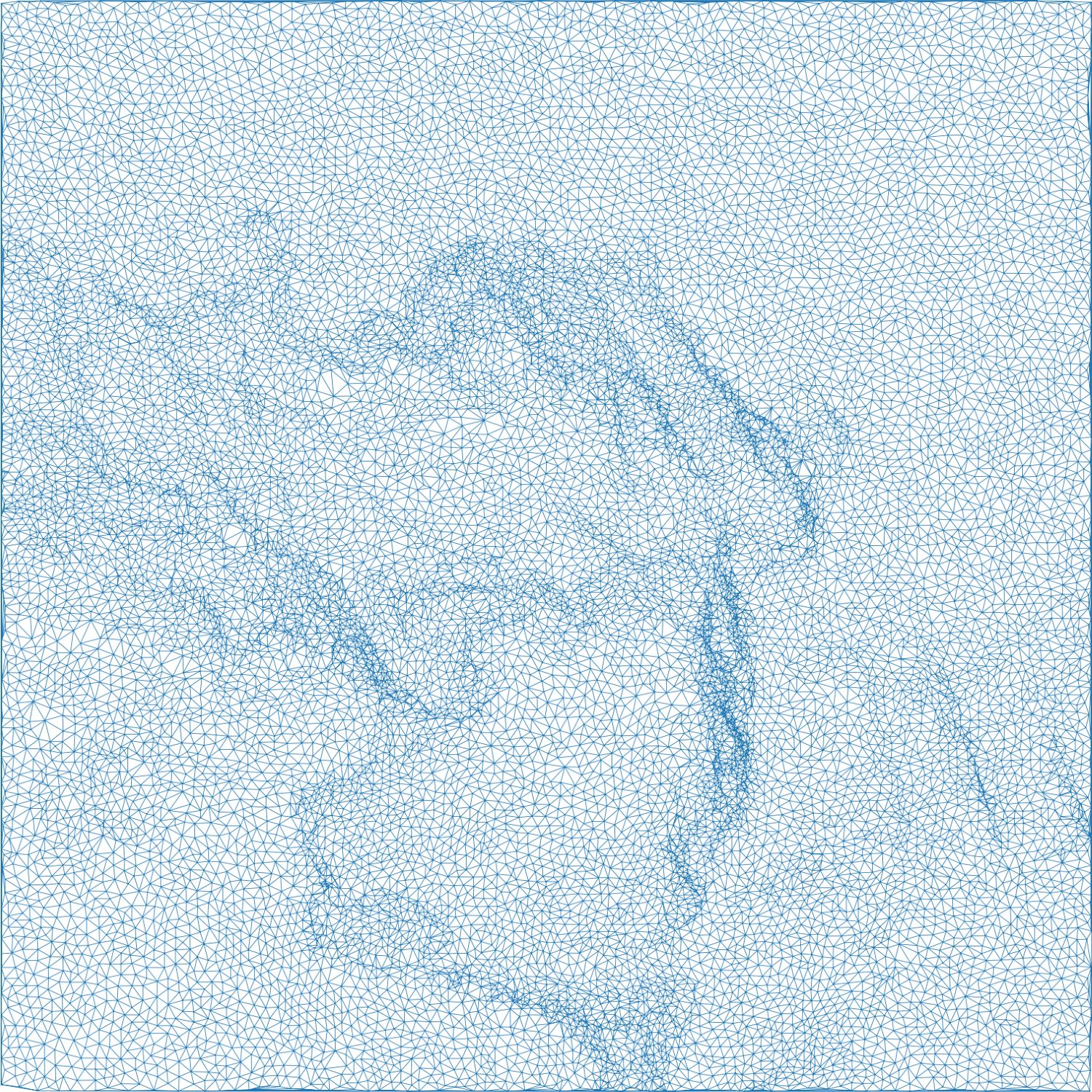}
        \subcaption{TIN - FPS down-sampling}
    \end{subfigure}
    \hfill
    \begin{subfigure}{0.49\columnwidth}
        \centering
        \includegraphics[width=\textwidth]{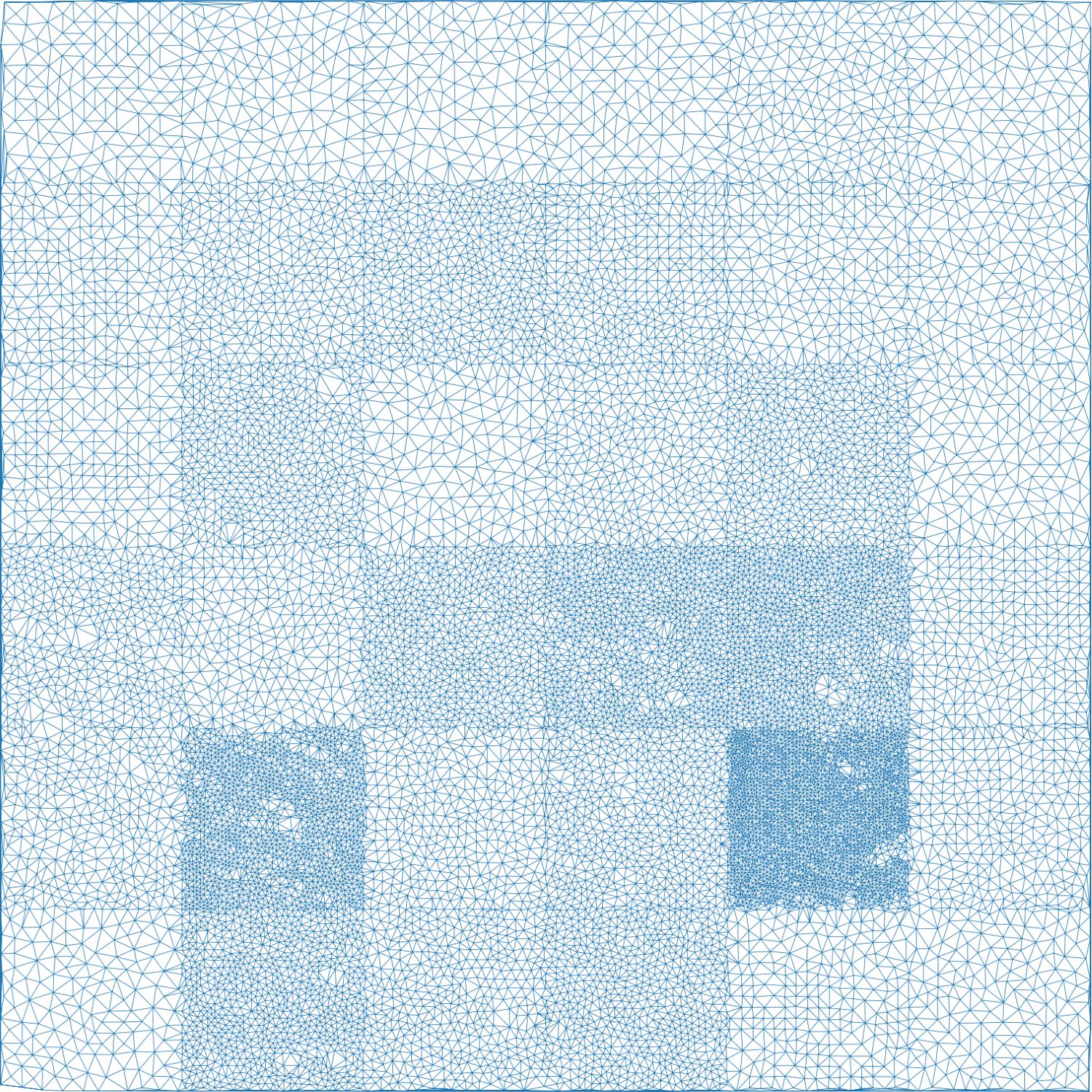}
        \subcaption{TIN - PFPS down-sampling}
    \end{subfigure}
    \caption{Comparison between TINs generated by point clouds down-sampled by different methods.}\label{fig:downsampling}
    \Description{Comparison between TINs generated by point clouds down-sampled by different methods.}
\end{figure}

Different point cloud down-sampling methods cater to specific needs. For example, uniform down-sampling reduces point numbers for visualization purposes, while our curvature-based down-sampling method preserves topographical information. This section compares these down-sampling methods' performance on the dataset, with results shown in Figure~\ref{fig:downsampling}.

Analyzing the results reveals that the initial point cloud, characterized by non-uniform distribution and `scanlines' of Light Detection and Ranging (LiDAR), necessitates effective and accurate down-sampling to generate a suitable TIN for the proposed scale-space method. Random down-sampling, while straightforward, fails to maintain a stable number of adjacent vertices in the TIN. Voxel down-sampling closely resembles the grid-based method but inefficiently allocates memory and computational resources to flat areas, which are of little interest to terrain topological analysis. Furthest Point Sampling (FPS) down-sampling produces a well-balanced TIN regarding adjacent vertex numbers and point density but shares the inefficiencies of voxel down-sampling and leads to point clustering in steep terrain regions.

To overcome these downsides, we modified FPS down-sampling to be patch-based and prioritized points based on $xy$-coordinates. Patches with higher estimated curvature values contain more terrain topology information, resulting in denser point sampling. Consequently, Patch-based Furthest Point Sampling (PFPS) down-sampling generates a well-balanced TIN suitable for our TIN-based scale-space method.

\section{More experiment results}\label{appendix:prfresults}
\begin{figure}[!htbp]
    \centering
    \begin{subfigure}{0.95\columnwidth}
        \centering
        \includegraphics[width=\textwidth]{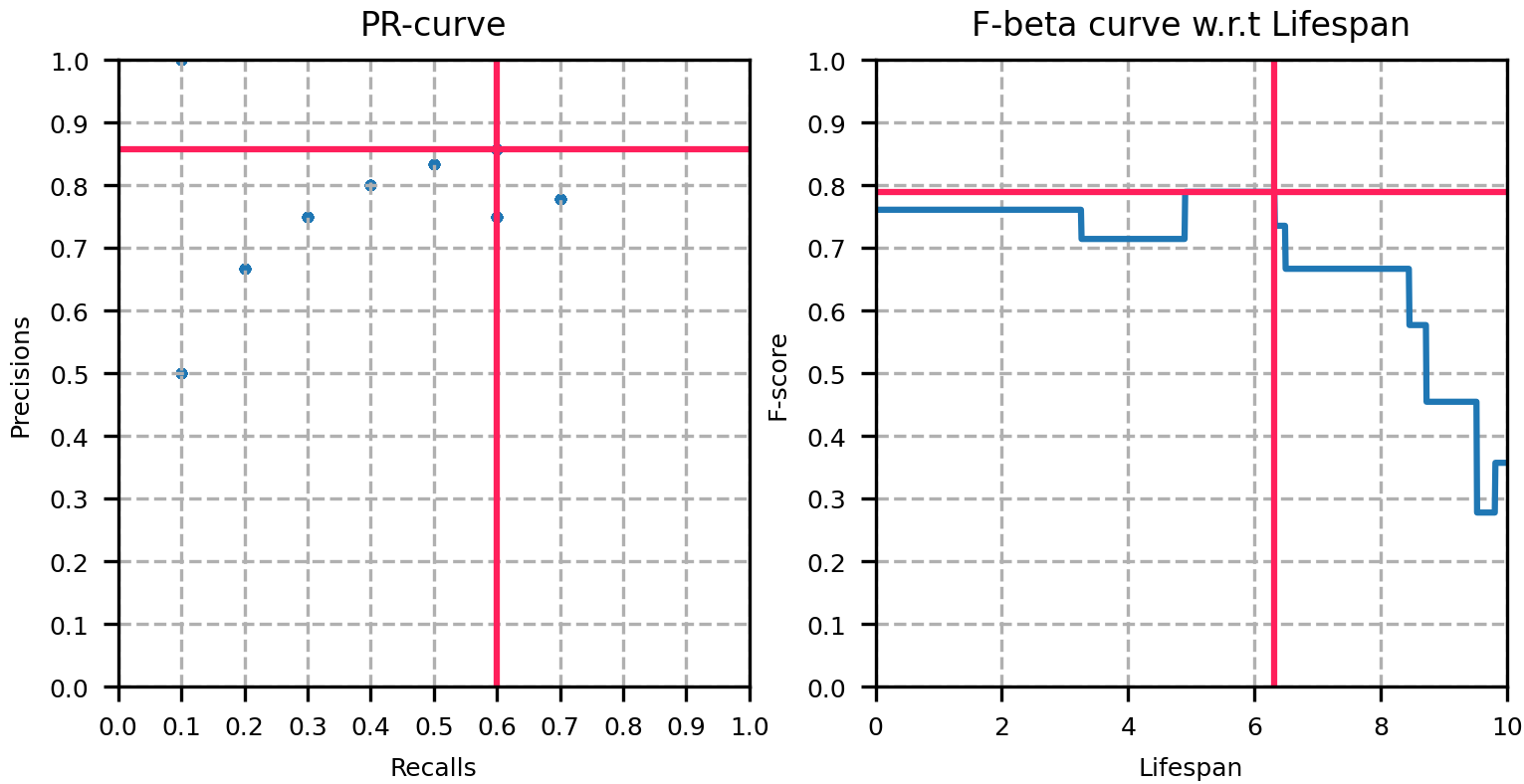}
        \subcaption{Grid-based method - best $F_{\beta} = 0.79$}
    \end{subfigure}
    \begin{subfigure}{0.95\columnwidth}
        \centering
        \includegraphics[width=\textwidth]{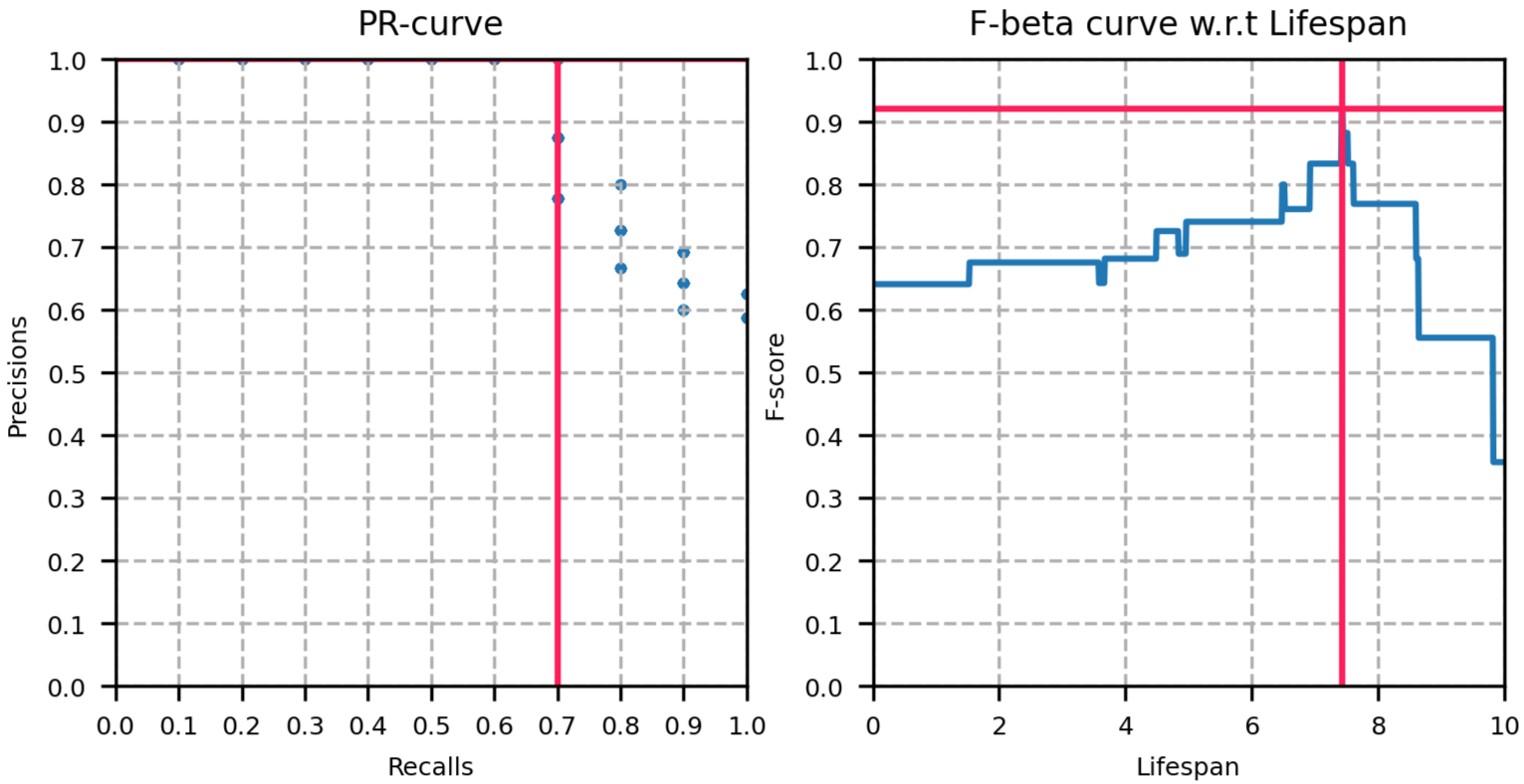}
        \subcaption{TIN-based method - best $F_{\beta} = 0.92$}
    \end{subfigure}
    \caption{PR curve and $F_{\beta}$ score graph of the grid-based and TIN-based methods on the Reichenburg suburban region.}
    \label{fig:prf_reichen}
    \Description{PR curve and $F_{\beta}$ score graph of the grid-based and TIN-based methods on the Reichenburg suburban region.}
\end{figure}

Figure~\ref{fig:prf_reichen} presents the PR curve and $F_{\beta}$ score graph for grid-based and TIN-based methods applied to the Reichenburg suburban region. Here, the TIN method attains a superior $F_{\beta}$ score, and its high recall rate shows its ability to identify more initial critical points than the grid-based method, benefiting from the applied down-sampling strategy.

\begin{figure}[!htbp]
    \centering
    \begin{subfigure}{\linewidth}
        \centering
        \includegraphics[width=\textwidth]{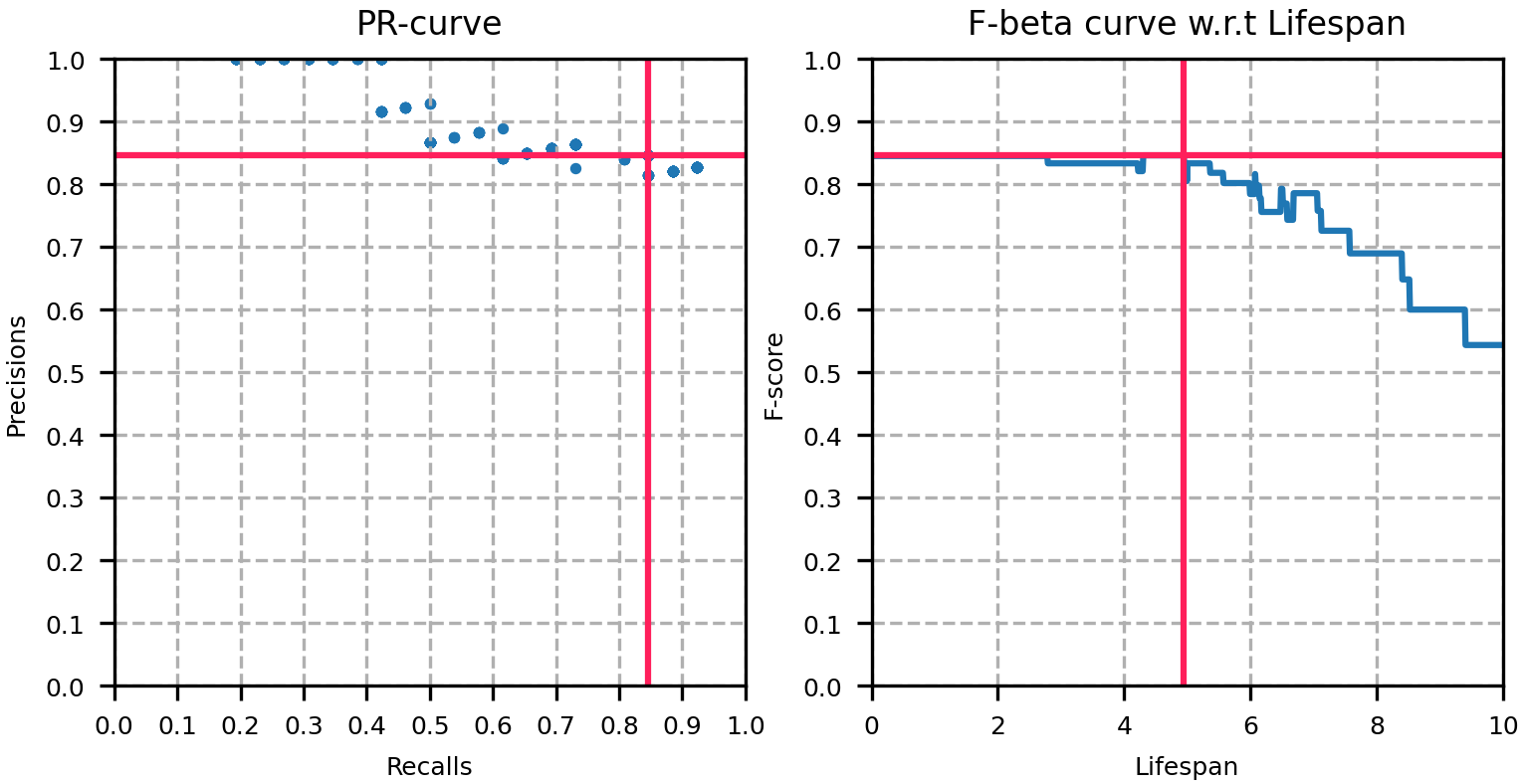}
        \subcaption{Grid-based method - best $F_{\beta} = 0.85$}
    \end{subfigure}
    \begin{subfigure}{\linewidth}
        \centering
        \includegraphics[width=\textwidth]{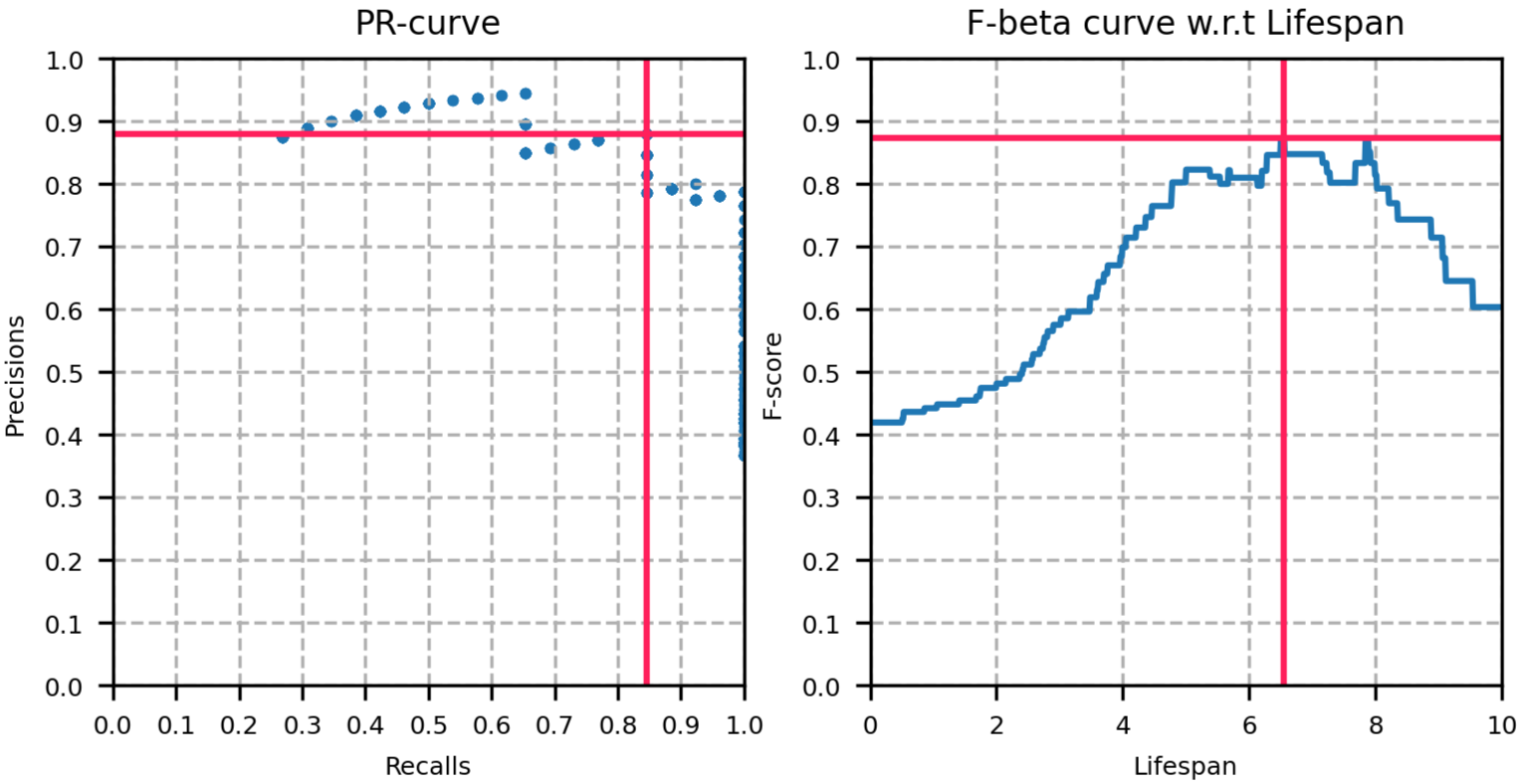}
        \subcaption{TIN-based method - best $F_{\beta} = 0.87$}
    \end{subfigure}
    \caption{PR curve and $F_{\beta}$ score graph of the grid-based and TIN-based methods on the Sörenberg region.}
    \label{fig:prf_soren}
    \Description{PR curve and $F_{\beta}$ score graph of the grid-based and TIN-based methods on the Sörenberg region.}
\end{figure}

Figure~\ref{fig:prf_soren} illustrates the PR curve and $F_{\beta}$ score graph of both methods applied to the Sörenberg mountainous region. The TIN-based method outperforms the grid-based method slightly in terms of the $F_{\beta}$ score, with a higher recall rate. Due to the high resolution sampled in the mountainous region, the TIN method identifies more critical points than the spot height selection task needs. We argue that it is due to the lack of constraint on the spot height placement in our post-processing stage.

\section{Gridded DEMs and TINs at different resolutions}\label{appendix:resolution}
\begin{figure}[H]
    \centering
    \begin{subfigure}{0.49\columnwidth}
        \centering
        \includegraphics[width=\textwidth]{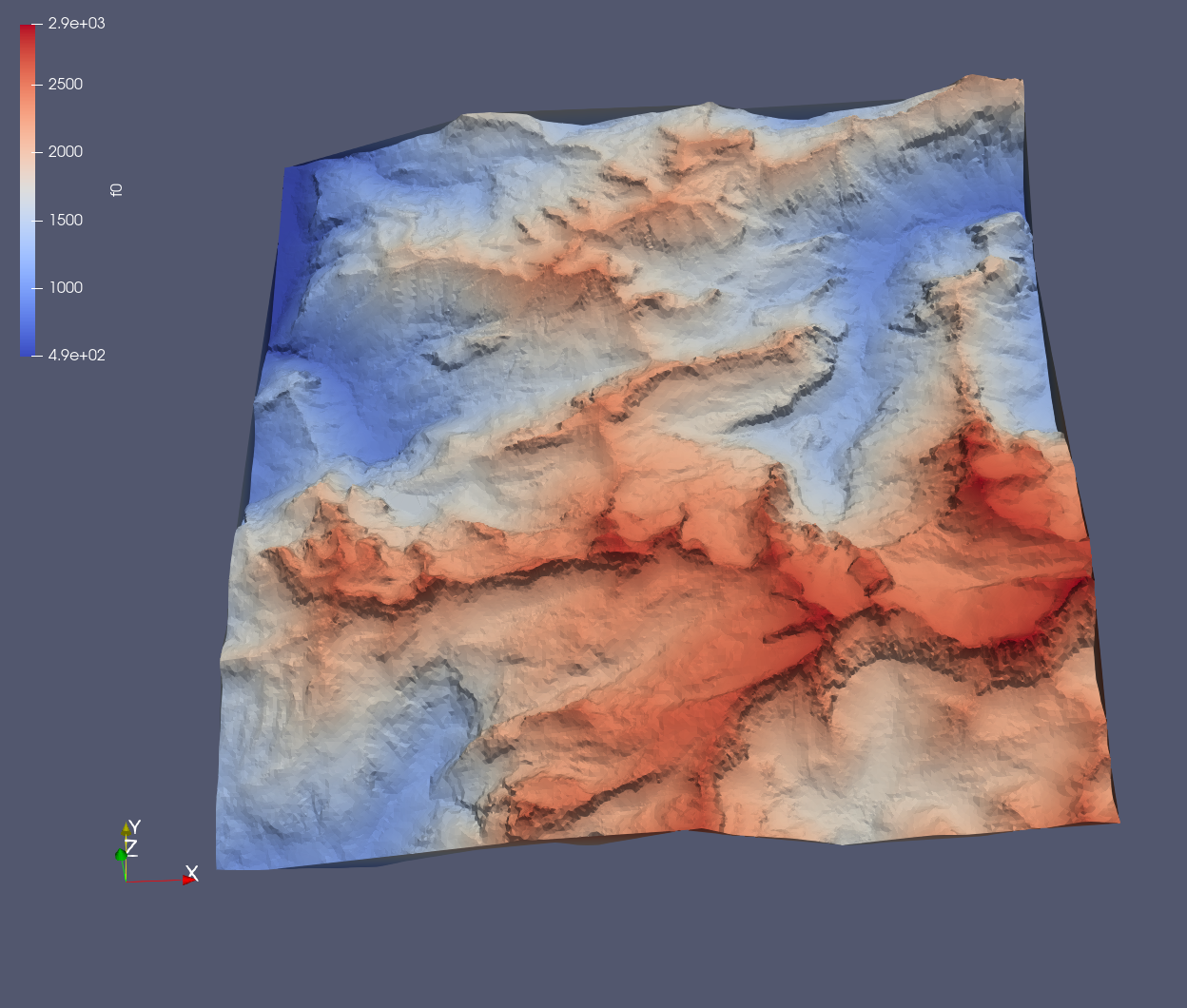}
        \subcaption{TIN - $40$ resolution.}
    \end{subfigure}
    \hfill
    \begin{subfigure}{0.49\columnwidth}
        \centering
        \includegraphics[width=\textwidth]{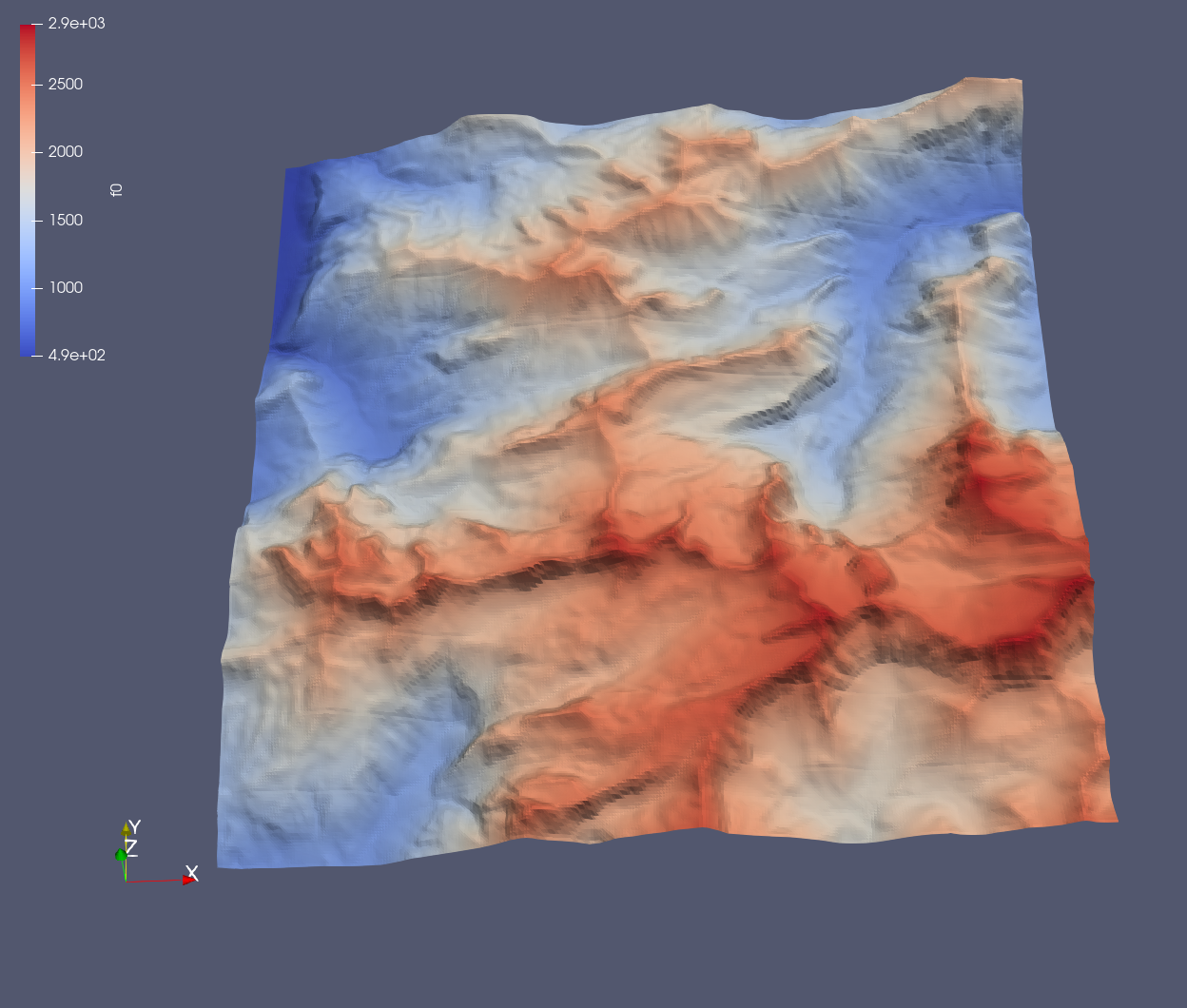}
        \subcaption{Gridded DEM - $40$ resolution.}
    \end{subfigure}
    \begin{subfigure}{0.49\columnwidth}
        \centering
        \includegraphics[width=\textwidth]{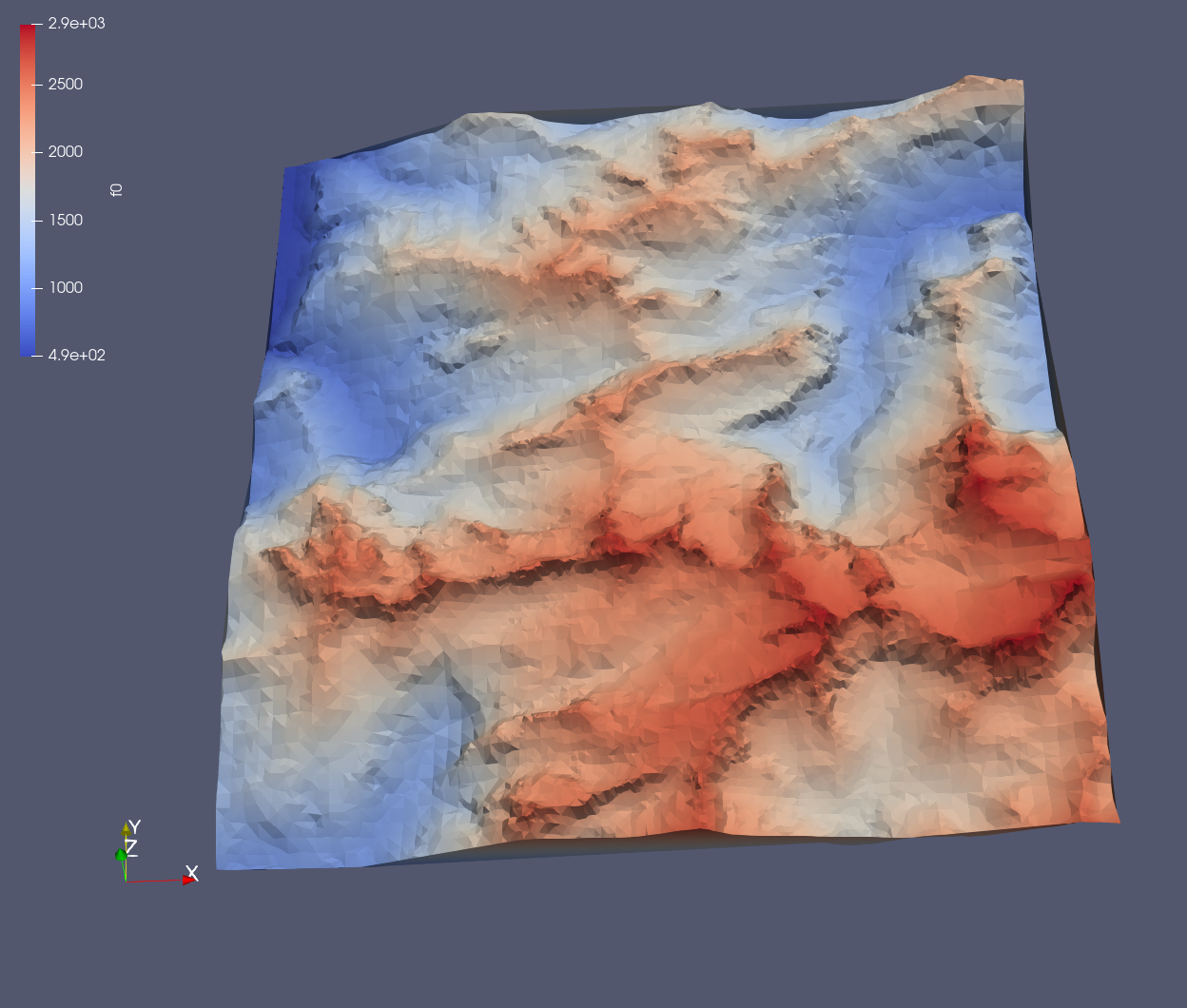}
        \subcaption{TIN - $80$ resolution.}
    \end{subfigure}
    \hfill
    \begin{subfigure}{0.49\columnwidth}
        \centering
        \includegraphics[width=\textwidth]{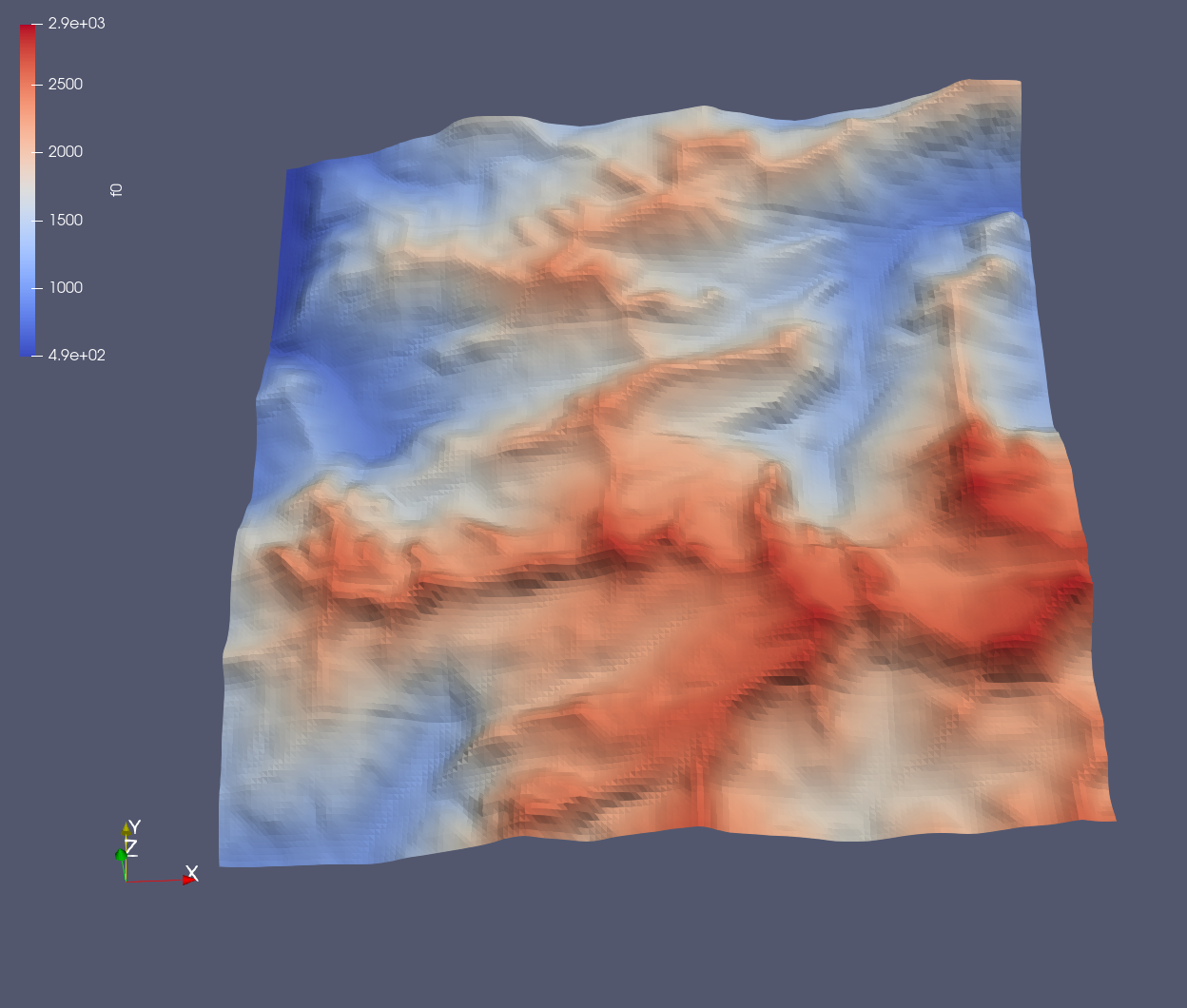}
        \subcaption{Gridded DEM - $80$ resolution.}
    \end{subfigure}
    \begin{subfigure}{0.49\columnwidth}
        \centering
        \includegraphics[width=\textwidth]{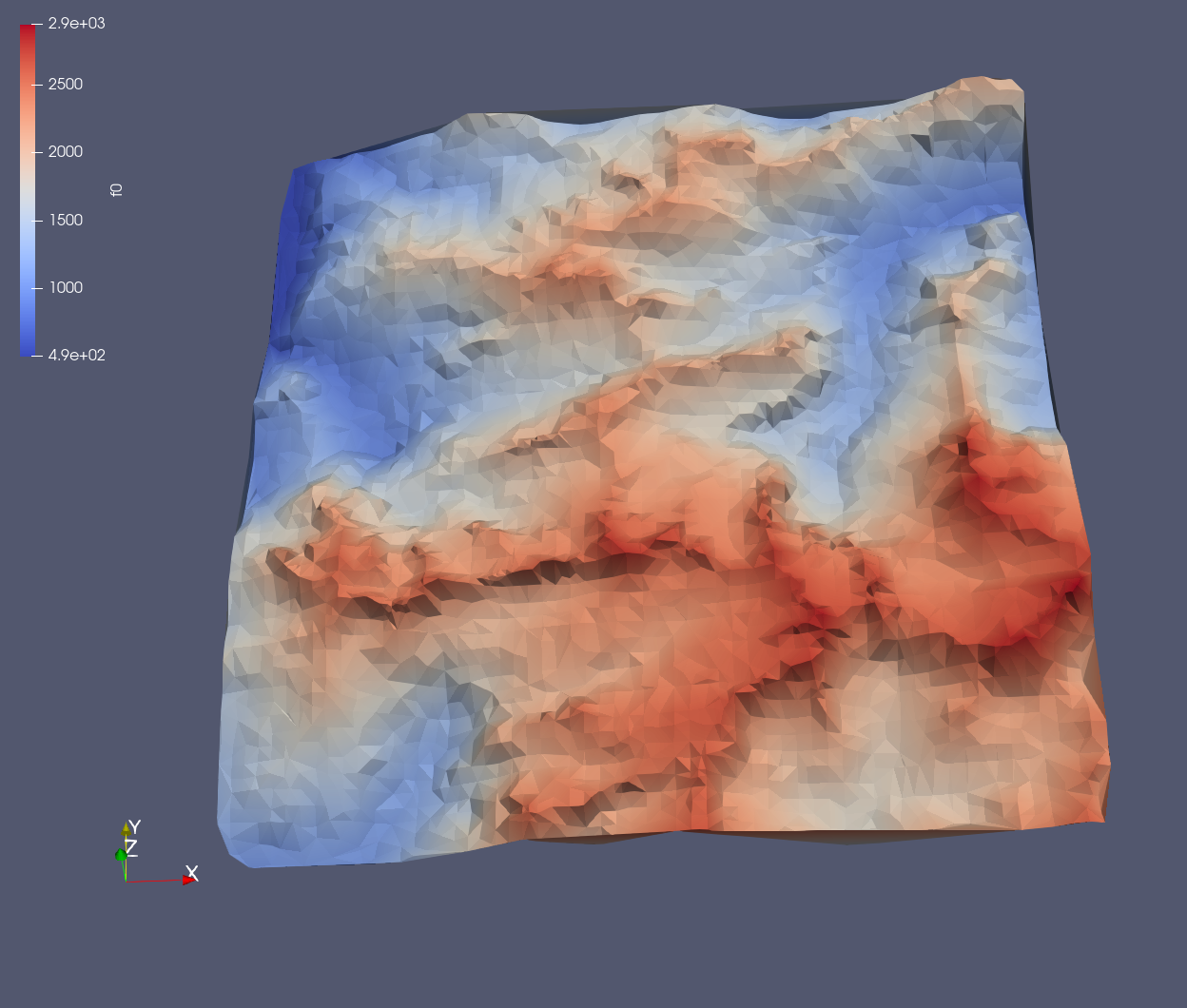}
        \subcaption{TIN - $160$ resolution.}
    \end{subfigure}
    \hfill
    \begin{subfigure}{0.49\columnwidth}
        \centering
        \includegraphics[width=\textwidth]{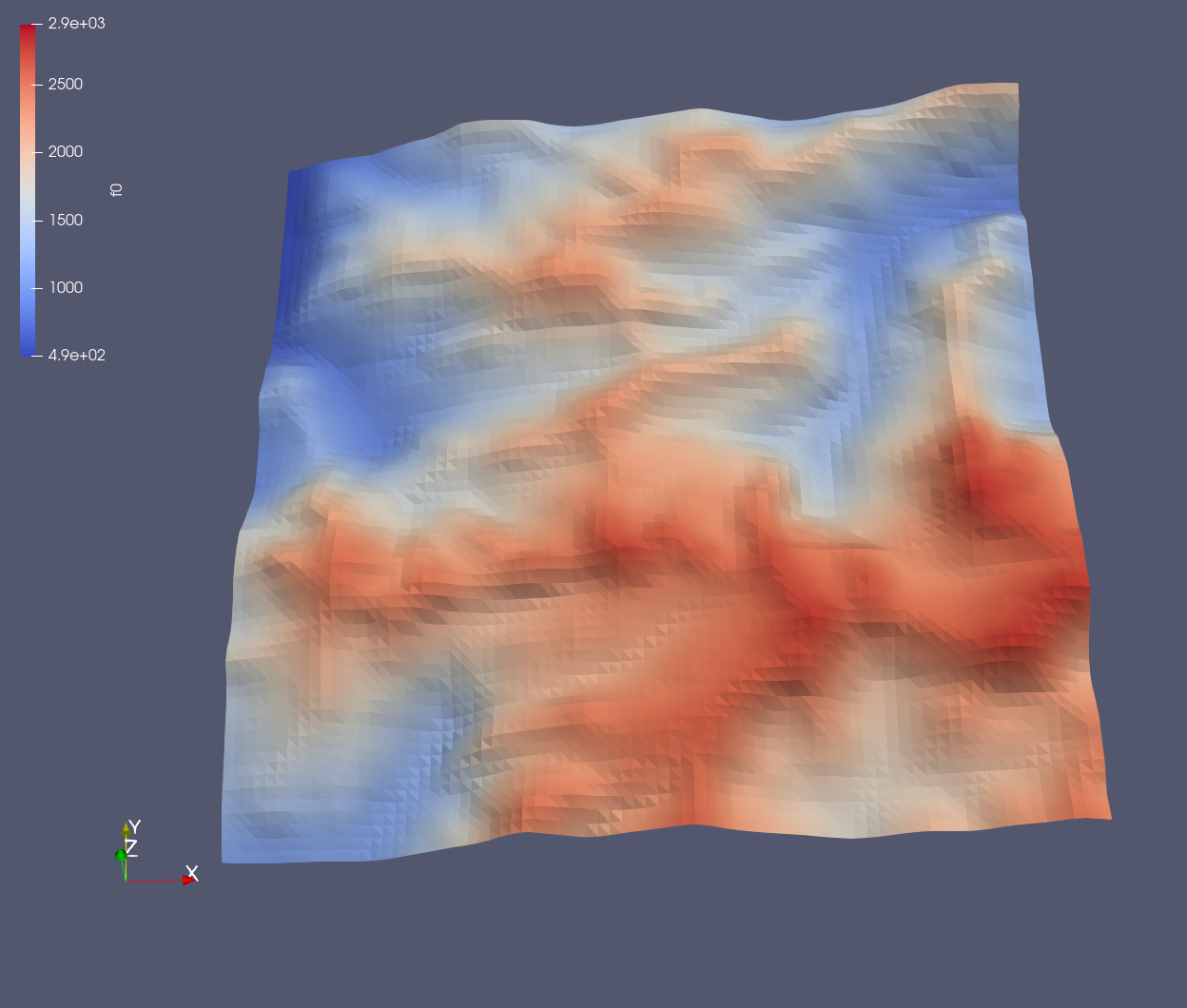}
        \subcaption{Gridded DEM - $160$ resolution.}
    \end{subfigure}
    \begin{subfigure}{0.49\columnwidth}
        \centering
        \includegraphics[width=\textwidth]{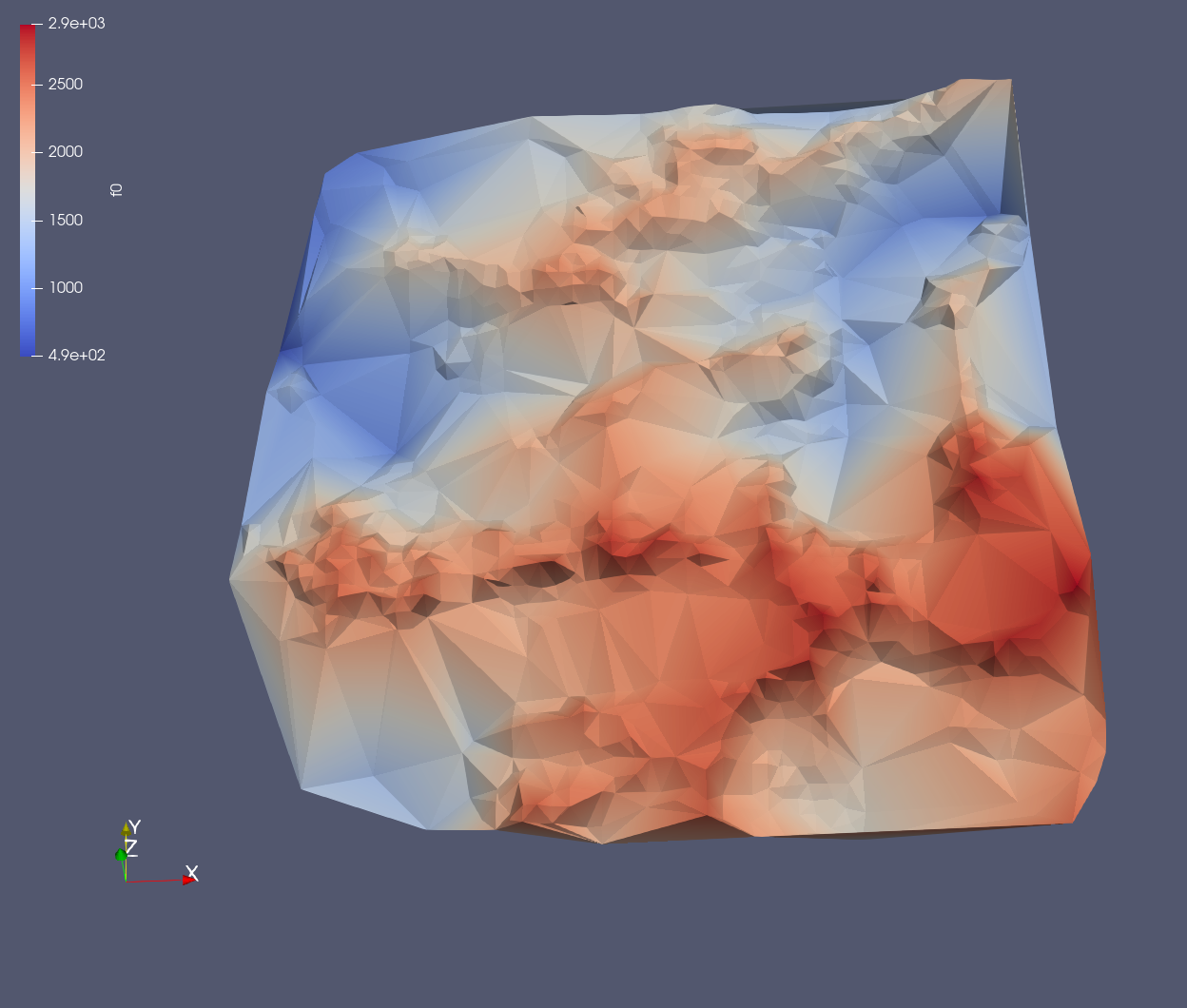}
        \subcaption{TIN - $300$ resolution.}
    \end{subfigure}
    \hfill
    \begin{subfigure}{0.49\columnwidth}
        \centering
        \includegraphics[width=\textwidth]{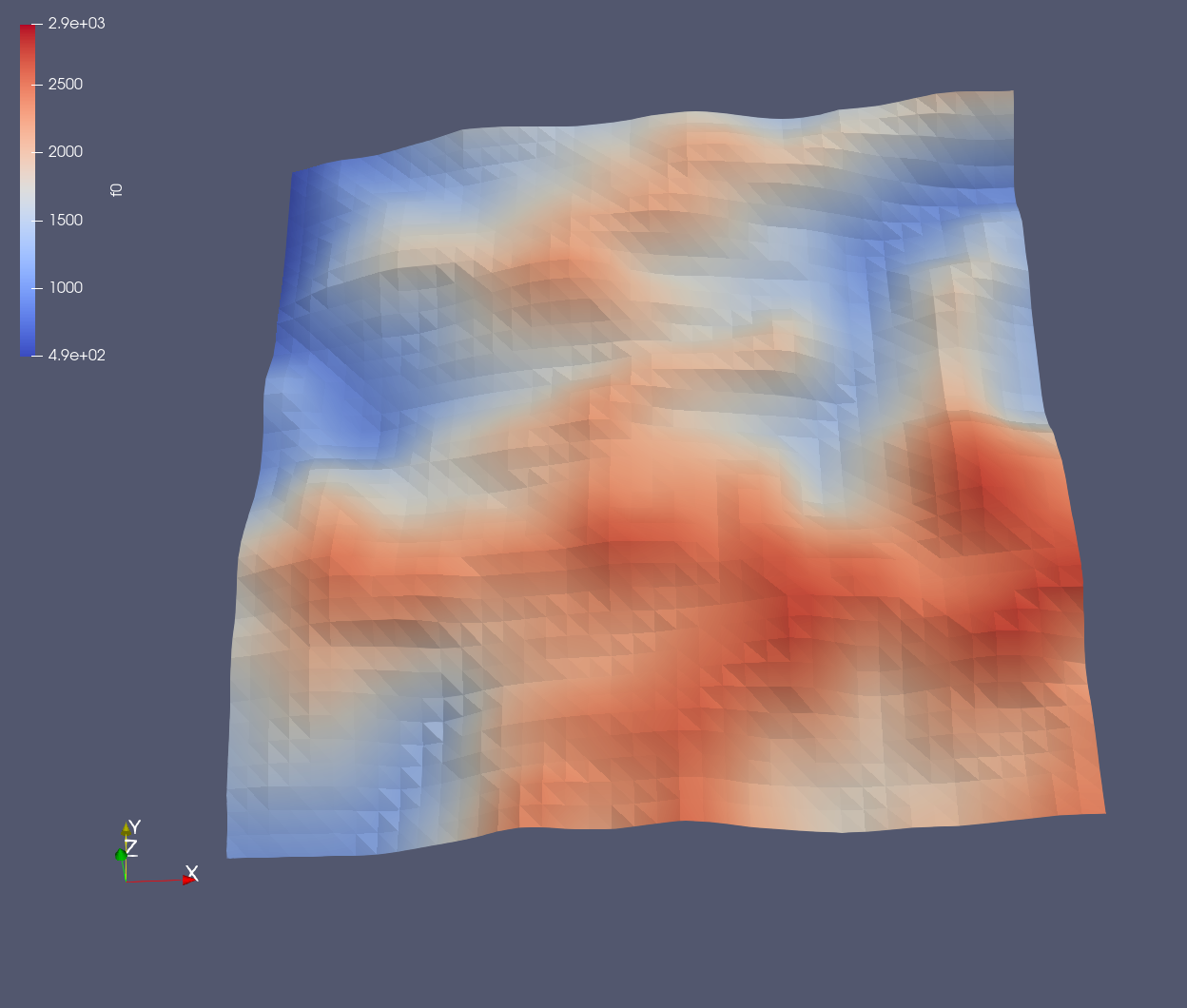}
        \subcaption{Gridded DEM - $300$ resolution.}
    \end{subfigure}

    \caption{Comparison between TINs and gridded DEMs at different resolutions. The resolution value denotes the grid cell size of the gridded DEM.}
    \label{fig:tin_grid_res}
    \Description{Comparison between TINs and gridded DEMs at different resolutions. The resolution value denotes the grid cell size of the gridded DEM.}
\end{figure}

Figure~\ref{fig:tin_grid_res} depicts the gridded DEMs and TINs of the Bannelpsee region at varying resolutions. Unlike the gridded DEM's smoothing-like result, TINs of different resolutions retain topographically significant regions, such as summits, ridges, and valleys. This preservation owes itself to our preprocessing stage's down-sampling strategy, facilitating the high recall rate of summit and peak identification in the spot height selection task, as evidenced by Table~\ref{table:result} and resolution robustness analysis in Figure~\ref{fig:resolution}.

\end{document}